\documentclass{jfm}
\usepackage{graphicx}
\usepackage{floatrow}
\usepackage{epstopdf, epsfig}
\usepackage{upgreek}
\usepackage{subfigure}
\usepackage{amsmath}
\usepackage{amssymb}
\usepackage{mathrsfs}  
\usepackage{nomencl}
\usepackage{setspace}
\usepackage[round]{natbib}
\usepackage[usenames,dvipsnames]{xcolor}
\usepackage{tikz}
\usetikzlibrary{shapes}
\usepackage{MnSymbol}
\usepackage[normalem]{ulem}
\usepackage[ruled,vlined]{algorithm2e}
\usepackage{wrapfig}
\usepackage{colortbl}
 \usepackage{hyperref}
\hypersetup{
	colorlinks=true, 
	linktoc=all,     
	linkcolor=blue,  
	citecolor=blue,
	urlcolor = blue
}

\DeclareRobustCommand\sampleline[1]{%
  \tikz\draw[#1] (0,0) (0,\the\dimexpr\fontdimen22\textfont2\relax)
  -- (2em,\the\dimexpr\fontdimen22\textfont2\relax);%
}

\def\m1line{\vrule width3pt height2.5pt depth -2pt}

\def\bdot{\raise.2em\hbox to .15em{.}}

\usepackage{todonotes}
\def\outline#1{\textcolor{Gray}{}}
	\long\def\comment#1{}

\allowdisplaybreaks[1]

\shorttitle{Difficulty of turbulence reconstruction from wall observations} 
\shortauthor{Q. Wang, M. Wang \& T. A. Zaki} 

\title{From wall observations to turbulence: \\ The difficulty of flow reconstruction}

\author{Qi Wang, Mengze Wang \and Tamer A. Zaki\corresp{\email{t.zaki@jhu.edu}}}

\affiliation{Department of Mechanical Engineering, Johns Hopkins University, Baltimore, MD 21218, USA}
\begin{document}

\maketitle


\begin{abstract}
Estimation of the initial state of turbulent channel flow from spatially and temporally resolved wall data is performed using adjoint-variational data assimilation.
The accuracy of the predicted flow deteriorates with distance from the wall, most precipitously across the buffer layer beyond which only large-scale structures are reconstructed.
To quantify the difficulty of the state estimation, the Hessian of the associated cost function is evaluated at the true solution.
The forward-adjoint duality is exploited to efficiently compute the Hessian matrix from the ensemble-averaged cross-correlation of the adjoint fields due to impulses at the sensing locations and times.
Characteristics of the Hessian are examined when observations correspond to the streamwise or spanwise wall shear stress or wall pressure.
Most of the eigenmodes decay beyond the buffer layer, thus demonstrating weak sensitivity of wall observations to the turbulence in the bulk. 
However, when the measurement time $t_m^+ \gtrsim 20$, some streamwise-elongated Hessian eigenfunctions remain finite in the outer flow, which corresponds to the sensitivity of wall observations to outer large-scale motions. 
Furthermore, we report statistics of the adjoint-field kinetic-energy budget at long times, which are distinctly different from those of the forward model.  
One notable difference is the high concentration of energy within the buffer layer and its narrower support. 
A large and exponentially amplifying adjoint field in that region leads to large gradients of the cost function, smaller step size in the gradient-based optimization, and difficulty to achieve convergence especially elsewhere in the domain where the gradients are comparatively small. 
\end{abstract}

%

\section{Introduction}

In turbulent channel flow, the walls play an important role in the generation of vorticity.  
Despite this role, whether wall observations of the stresses can be used in data assimilation to accurately predict the initial state of the flow and its evolution are challenging questions.  
The difficulty is primarily due to the non-linear, chaotic nature of turbulence: small deviations in the initial state grow exponentially in time \citep{Deissler1986chaotic} and, conversely, small errors in the observations can obscure the reconstruction of the flow.
Previous studies have demonstrated that the near-wall turbulence and only the outer large-scale structures can be reconstructed from wall signals. 
In this work, we adopt an adjoint-variational approach to quantify the domain of dependence of wall sensors, and demystify the difficulty of flow reconstruction from wall observations.

Previous attempts to estimate the state of wall-bounded turbulence from limited observations can be separated into two general classes: filters and smoothers \citep[e.g.][]{colburn2011state,Suzuki2012,mons2016reconstruction,wang_hasegawa_zaki_2019}.
In filtering techniques, the state is adjusted to best match the most recently available observations.
On the other hand, smoothers take into account the entire observation history.

Extended \citep{hoepffner2005state} and ensemble \citep{chevalier2006state} Kalman filters were adopted to estimate turbulent channel flow from wall observations, and in both cases accuracy deteriorates with distance from the wall.
\citet{hasegawa2016estimation} adopted linear stochastic estimation (LSE) to interpret wall observations at friction Reynolds number ($Re_\tau = 100$). Accuracy of their predicted flow fields was commensurate with earlier efforts that used Kalman filtering techniques:  
The estimation was only successful in the region $y^+ \lesssim 20$, where $+$ indicates scaling in viscous wall units, and the error in the estimated field amplified beyond that height.  
At higher friction Reynolds number ($Re_\tau = 1000$-$5000$), LSE was able to also capture the large-scale structures in the outer layer \citep{encinar2019logarithmic}.
It is important to note that LSE relies on prior knowledge of the correlation between wall observations and the flow, which may not be available.  In addition, LSE and also Kalman filtering techniques do not satisfy the Navier-Stokes equations.

In contrast to filters, smoothers satisfy the governing equations and attempt to reproduce the observations over the entire time horizon during the forward evolution of the flow.  
The problem is formulated as a nonlinear optimization: An initial state is sought to minimize a cost function that is proportional to the difference between predicted and available observations.  Both ensemble-variational  \citep{mons2019kriging,buchta2021envar} and adjoint-variational (4DVar) \citep{Dimet1986_4dvar,Li2020} methods can be adopted for the minimization procedure.  In the latter approach, which will be adopted herein, deviations from the observations appear in the adjoint system as a forcing term; The adjoint variable at the initial time provides the gradient of the cost function with respect to the initial state, and is used to adjust the initial condition. 
Adjoint variational methods are popular in weather prediction \citep{kleist2015osse1, kleist2015osse2}, nonlinear stability analysis \citep{Schmid2007nonmodal,Kerswell2018nonlinear}, and optimal flow control \citep{Luchini2014adjoint}.
In wall turbulence, \citet{bewley2004skin} used 4DVar to reconstruct the initial state of channel flow from observations of wall friction and pressure, at modest friction Reynolds number $Re_{\tau} = 100$.
The estimation was accurate near the wall, but was nearly uncorrelated with the true initial flow state in the channel center.
As the Reynolds number is increased, the outer large-scale structures can also be decoded from wall observations using 4DVar \citep{mengze2021}.


A measure of the difficulty of reconstructing the turbulent state from wall data is desirable.  
In variational methods, since the optimal initial condition minimizes the cost function, the local gradient vanishes.
Therefore the second-order sensitivity of the cost function with respect to the initial condition, or its Hessian matrix, characterizes the local geometric properties of the optimization problem.
While the Hessian matrix has been referenced in earlier studies, it has never been evaluated for the reconstruction of channel-flow turbulence from wall data\textemdash a gap that we address herein. 
\cite{papadimitriou2008direct} compared four approaches to compute the Hessian matrix in an aerodynamic inverse design problem.
The Newton method based on exact Hessian matrices outperforms other gradient-based approaches such as steepest descent or L-BFGS. 
Furthermore, the eigen-decomposition of the Hessian matrix can be used for sensitivity analysis, and is also a powerful tool for uncertainty quantification.
For example, \citet{kalmikov2014hessian} estimated the large-scale ocean state using adjoint method and evaluated the Hessian matrix using algorithmic differentiation (AD).  Using the leading eigenvectors of the Hessian, they analyzed the influence of uncertainties in observations on estimation of oceanographic target quantities.


The adjoint-variational approach for state estimation is reviewed briefly in \S\ref{Sec:AdjointOpt}, and sample flow reconstructions in turbulent channel flow are presented.  
Using the forward-adjoint duality relation, we formulate an approach for computing the Hessian matrix at the true initial flow state in \S\ref{Sec:Linearized}.  
The adjoint fields associated with wall observations are reported in \S\ref{Sec:AdjointImpulse}, and interpreted as the dependence of wall data on the flow at different times from the observations. 
The adjoint fields are then used to construct the Hessian matrix.
Eigen-analysis for the Hessian is presented in \S\ref{Sec:EigenAnalysis}, and used to explain the sensitivity of the wall data to various wall-normal locations in state estimation. 
We proceed to evaluate the sensitivity of wall observations to the most energetic flow structures that are obtained from a proper orthogonal decomposition (POD) of turbulent channel flow. 
Finally, statistical behaviour of the adjoint-field at long times is discussed in \S\ref{Sec:AdjointAsymptotic}.

\section{Methods}
\label{Sec:Methods}

The flow configuration of interest is statistically stationary turbulent channel flow. The reference length $h$ is the channel half height and the reference velocity is the bulk value $\mathcal{U}$.  
The flow is governed by the non-dimensional incompressible Navier-Stokes equations,
\begin{subequations}
\label{Eqn:NS}
    \begin{eqnarray}
        \frac{\partial \mathbf{U}}{\partial t} + \left(\mathbf{U} \cdot \boldsymbol{\nabla}\right) \mathbf{U} &=& -\boldsymbol{\nabla} P + \frac{1}{Re}\nabla^2 \mathbf{U}, \\
        \boldsymbol{\nabla} \cdot \mathbf{U} &=& 0,  \\
        \left.\mathbf{U}\right\rvert_{t=0} &=& \mathbf{U}_0, 
    \end{eqnarray}
\end{subequations}
where $Re \equiv \mathcal{U} h / \nu $ is the bulk Reynolds number and $\nu$ is the kinematic viscosity.
The spatially and temporally dependent velocity and pressure are denoted $\mathbf{U}$ and $P$, and $\mathbf{U}_0$ is the initial state which is the target of reconstruction. 
The flow is assumed to be periodic in the streamwise ($x$) and spanwise ($z$) directions, with no-slip conditions at the bottom and top walls $y = \{0,2\}$.
The governing equations (\ref{Eqn:NS}) are solved using a fractional step method with a volume-flux formulation \citep{Rosenfeld_1991jcp}. 
The diffusion terms are discretized implicitly in time with Crank-Nicolson scheme while the nonlinear advection terms are treated explicitly using Adams-Bashforth.
The pressure Poisson equation is solved by Fourier transforms in the periodic streamwise and spanwise directions, followed by a tridiagonal solver in the wall-normal direction. 
The numerical algorithm has been used extensively for direct numerical simulations of transitional \citep{zaki2013streaks} and turbulent flows \citep{lee2017signature}.

The majority of the results are focused on Reynolds numbers $Re = 2{,}800$ and $10{,}935$; the corresponding friction Reynolds numbers are $Re_\tau \equiv u_\tau h / \nu =180$ and $Re_\tau=590$, where $u_\tau \equiv \sqrt{\nu \left(d\overline{U}/dy\right)_{wall}}$ is the friction velocity evaluated from the mean wall shear stress, and overbar denotes averaging in the homogeneous spatial directions and in time.  
We adopted a Cartesian grid with uniform spacing in both the streamwise and spanwise directions and hyperbolic stretching in the wall-normal coordinate.
The dimensions of the computational domains and grid resolutions are reported in table \ref{TABLE:Resolution}.  
The domain sizes and resolutions for additional Reynolds numbers that are referenced in the text are also provided in the table.

Independent reference simulations were performed to generate the observation data that are then adopted in the state estimation procedure.  The reference flow fields are therefore the hidden truth that will be used to quantify the accuracy of the estimated fields.  

\begin{table}
	\centering
	\begin{tabular}{cccccccccccc}
	\hline
	\multicolumn{2}{c}{Parameters} & \multicolumn{3}{c}{Domain size}  & \multicolumn{3}{c}{Grid points} & \multicolumn{4}{c}{Grid resolution} \\
		 $Re_{\tau}$ & $Re$ & $L_x/h$ & $L_y/h$ & $L_z/h$ & $nx$ & $ny$ & $nz$ &  $\Delta x^+$ & $\Delta z^+$ & $\Delta y_{max}^+$ &  $\Delta y_{min}^+$ \\ 
	\hline
		~~100   & ~1{,}429 &  4$\pi$   & 2  & 2$\pi$   & 128  & 128   & 128  & 9.8   & 4.9   & 2.2~  & 0.74  \\
\rowcolor{blue!10}		
        ~~180   & ~2{,}800 &  4$\pi$   & 2  & 2$\pi$   & 384  & 256   & 320  & 5.9   & 3.5   & 2.95  & 0.20  \\
		~~392   & ~6{,}875 &  2$\pi$   & 2  & $\pi$   & 256  & 320   & 192  & 9.6   & 6.4   & 5.1~  & 0.34  \\
\rowcolor{blue!10}		
        ~~590   & 10{,}935 &  2$\pi$   & 2  & $\pi$   & 384  & 384   & 384  & 9.6   & 4.8   & 6.5~  & 0.44  \\ 
        1{,}000 & 20{,}000 &  2$\pi$   & 2  & $\pi$   & 768  & 768   & 768  & 8.2   & 5.5   & 4.1~  & 0.37  \\ 
	\hline
	\end{tabular}
	\caption{Domain sizes and grid resolutions. Grid sizes are normalized by the viscous lengthscale, e.g.\,$\Delta x^+ \equiv \Delta x\,u_{\tau} / \nu$}.
	\label{TABLE:Resolution}
\end{table}

\subsection{Adjoint variational state estimation}
\label{Sec:AdjointOpt}

Starting from wall observations only, we seek to evaluate an estimate of the true initial state of the flow $\mathbf{U}_0$, which will be denoted $\tilde{\mathbf{U}}_0$.
Similar to $\mathbf{U}$, the estimated field $\tilde{\mathbf{U}}$ satisfies the Navier-Stokes equation (\ref{Eqn:NS}) which are referred to as the forward model. 
Errors in $\tilde{\mathbf{U}}_0$ result in deviation of the associated model predictions from available observations, and the difference is used in the definition of our cost function that we aim to minimize.
Throughout the present study, observations are only available at the top and bottom surfaces $S$ of the channel, and the final observation time is denoted $t_m$.
When observations are only available at one instant, $t_m$, the associated cost function for the streamwise wall stress is, 
\begin{equation}
\label{Eqn:CostFunctionNL}
        \mathcal{J}_u(\tilde{\mathbf{U}}_0;t_m) = \frac{1}{2S}\int_S \left( \frac{\partial\tilde{U}}{\partial y}-\frac{\partial U}{\partial y} \right)^2_{(\mathbf{x}_m,t_m)} dx_m dz_m,
\end{equation}
where $\mathbf{x}_m$ is the observation location.
Similarly, cost functions $\mathcal{J}_w$ and $\mathcal{J}_p$ are defined for deviations from instantaneous observations of spanwise wall stress $\left({\partial\tilde{W}}/{\partial y}-{\partial W}/{\partial y} \right)_{(\mathbf{x}_m,t_m)}$ and wall pressure $\left(\tilde{P}- P \right)_{(\mathbf{x}_m,t_m)}$.  

Results for observations of only one component of the stress or pressure at one time instance are the focus of this work, and will be discussed in detail in \S\ref{Sec:AdjointImpulse} and \S\ref{Sec:EigenAnalysis}.
In order to place that discussion in context, however, we start by examining the capacity for state estimation when all three components are available as a function of time during the assimilation window; this preliminary step is a summary of a recent detailed study by \citet{mengze2021}.  
The cost function for the collective observations over the entire time horizon is defined as, 
 \begin{eqnarray}
    J   &=& \frac{1}{Re^2} J_u + \frac{1}{Re^2} J_w + J_p \nonumber \\
        &=& \int_0^{t_m} \frac{1}{Re^2}\mathcal{J}_u(\bullet; t^{\prime}) + \frac{1}{Re^2}\mathcal{J}_w(\bullet;t^{\prime}) + \mathcal {J}_p(\bullet;t^{\prime})  dt^{\prime}.
    \label{Eqn:CostFunction_DA}    
\end{eqnarray}
In this form, the three contributions to the cost function are comparable in magnitude.
The gradient of the cost function with respect to the initial condition is obtained by solving the adjoint equations,
\begin{subequations}
    \label{Eqn:Adjoint_forcing}
    \begin{eqnarray}
	    \boldsymbol{\nabla} \cdot \mathbf{u}^* &=& -\frac{\partial J}{\partial \tilde{P}}, \\
	    	    \frac{\partial \mathbf{u^*}}{\partial \tau} + \left( \boldsymbol{\nabla} \tilde{\mathbf{U}} \right) \cdot \mathbf{u^*} - \left(\tilde{\mathbf{U}} \cdot \boldsymbol{\nabla} \right) \mathbf{u}^* &=&  \boldsymbol{\nabla} p^* + \frac{1}{Re} \nabla^2 \mathbf{u^*} + \frac{\partial J}{\partial \tilde{\mathbf{U}}}, 
    \end{eqnarray}
\end{subequations}
where $*$ denotes adjoint variables, and $\tau \equiv t_m-t$ is the reverse time over the duration of the assimilation window $t_m$. 
The source terms on the right side of the equations are due to the deviation between the estimated and true observations.
The gradient of the cost function with respect to the initial condition is obtained at the end of the adjoint computation, ${\partial J}/{\partial \tilde{\mathbf{U}}_0} = \mathbf{u}^*(\tau=t_m)$ \citep[for a detailed derivation, see][]{mengze2019discrete}.

The adjoint equations (\ref{Eqn:Adjoint_forcing}) are solved using the discrete adjoint approach, which provides a more accurate gradient of the cost function than the continuous adjoint counterpart \citep{Vishnampet2015}.
We adopted the limited-memory BFGS (L-BFGS) \citep{LBFGS} algorithm to utilize the gradient evaluated from different forward-adjoint loops to update our estimation of the initial state until we significantly reduce the cost function.
In summary, the procedure for minimizing the cost function (\ref{Eqn:CostFunctionNL}) is the following:
(i) Start with a guess of the initial condition $\tilde {\mathbf U}_0$, march the forward equations (\ref{Eqn:NS}) from $t=0$ to $t=t_m$, and store the instantaneous fields $\tilde {\mathbf U}(t)$;  (ii) Solve the adjoint equations (\ref{Eqn:Adjoint_forcing}) from $\tau = 0$ to $\tau = t_m$ and compute the gradient of the cost function; (iii) Update the estimated initial condition $\tilde {\mathbf U}_0$, and repeat procedures (i) and (ii) until convergence.
Note that the optimization process is computationally expensive, both in terms of storage cost and computational time. 
The former is caused by the storage of the forward flow fields $\tilde {\mathbf U}(t)$, and the latter is due to the hundreds of forward and adjoint simulations that are required for the minimization of the cost function.

Due to the exponential divergence of trajectories of both the forward and adjoint models, the choice of the assimilation time horizon is guided by the Lyapunov timescale.  In viscous units, that horizon is $t_m^+ = 50$, and fields were stored at every time step to perform adjoint simulations. 
At each Reynolds number, the initial guess of $\tilde{U}_0$ is obtained from linear stochastic estimation (LSE) \citep{Adrian1988LSE,encinar2019logarithmic} using wall measurements. This guess is then refined using one hundred L-BFGS iterations, each comprising a forward and an adjoint fully resolved simulations of turbulence over the time window of assimilation.
The convergence history for different Reynolds numbers are similar, and the cost function is reduced to less than $4\%$ of its initial value.

\begin{figure}
    \centering
    \includegraphics[width = 0.97\textwidth]{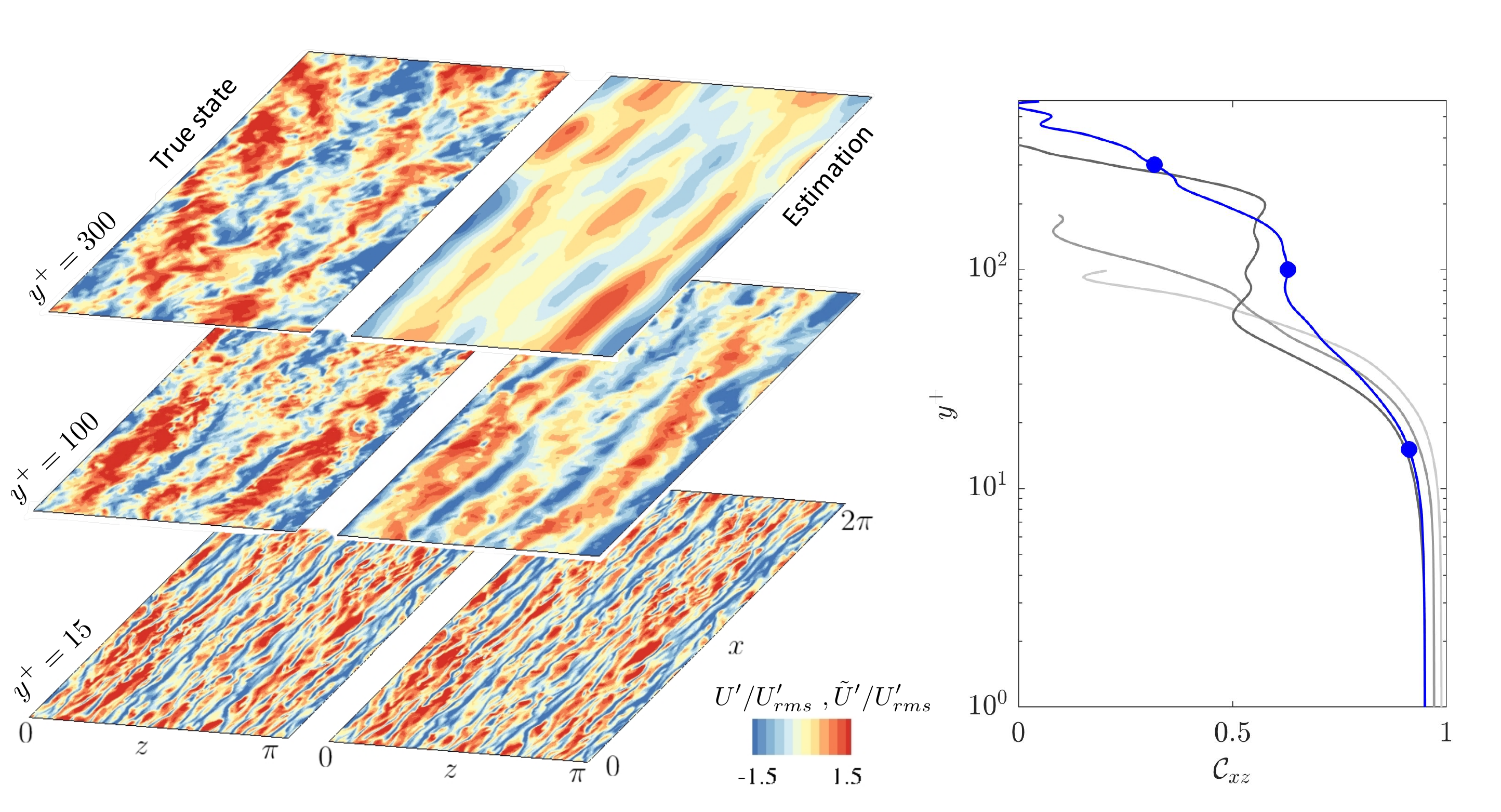}
    \caption{Left: Top views of streamwise velocity fluctuations in the true and estimated states at $t=t_m$ for $Re_{\tau} = 590$ at selected $y$ locations, $y^+ = \{15, 100, 300\}$. Right: Correlation coefficient between true and estimated streamwise velocities at $t=t_m$. Progressively darker lines corresponding to $Re_{\tau} = \{100,180,392\}$; thick blue curve corresponds to $Re_{\tau} = 590$;  filled circles mark the wall-normal locations of the left panels.  
    }
    \label{Fig:SampleReconstruction}
\end{figure}

A sample estimation of the turbulence from wall observations is provided in figure \ref{Fig:SampleReconstruction}.
Since the estimated flow progressively approaches the true state in time within the estimation window \citep[e.g.][]{mengze2019discrete}, we only report the predictions at $t=t_m$ when the estimation is most accurate.
The contours in figure \ref{Fig:SampleReconstruction}a show both the streamwise velocity fluctuations from the turbulence in the true state and in the reconstruction, for $Re_{\tau} = 590$. The two fields are nearly identical very near the wall, but precipitously deviate from one another from the buffer layer and into the bulk of the channel. 
The turbulence within the core of the channel is not accurately reconstructed, with the exception of the large-scale streaky structures; these outer energetic motions are known to modulate the near-wall region and the wall shear stress \citep{abe2004very,mathis2009large,hwang2016inner,you2019tbl}.
These characteristics of the estimated field are a summary of the findings by \citet{mengze2021}, and are consistent with earlier efforts \citep{bewley2004skin,hasegawa2016estimation,encinar2019logarithmic}.

The correlation coefficient $\mathcal C_{xz}$ between the true and estimated fields is reported in figure \ref{Fig:SampleReconstruction}$b$, 
\begin{equation}
		\label{Eqn:Coef}
		\mathcal C_{xz}  = \frac{ \langle \tilde{U}^{\prime} U^{\prime} \rangle_{xz} } 
		{ {\langle U^{\prime 2} \rangle_{xz}^{1/2}} { \langle \tilde{U}^{\prime 2} \rangle_{xz}^{1/2}} },
\end{equation}
where angle brackets denote averaging with respect to the marked dimensions and prime indicates fluctuations $U^{\prime} = U - \left\langle U\right\rangle_{xz}$. 
The correlation decays from the start of the buffer layer and the estimated field becomes essentially uncorrelated with the true state as we approach the channel center. 
We have repeated the analysis for Reynolds numbers $Re_\tau = \{100, 180, 392, 590\}$, and the findings remain qualitatively similar.  In effect, as the Reynolds number increases, wall observations can be decoded to accurately estimate the flow in a diminishingly small physical region near the wall and also the outer large-scale structures that are known to have a near-wall signature. 

\subsection{Hessian analysis}
\label{Sec:Linearized}

A fundamental difficulty in reconstructing the initial condition is due to the chaotic nature of the governing equations and their adjoint: just as infinitesimal deviation in the initial conditions leads to exponentially diverging trajectories in forward time, small mismatch between the estimated and true observations leads to exponentially diverging trajectories in the adjoint.
In terms of the variational state-estimation procedure, we can quantify this difficulty by examining the convergence behaviour when the estimated state $\tilde{\mathbf{U}}_0$ is infinitesimally close to the true solution $\mathbf{U}_0$.
We can then introduce the deviation field $\mathbf{u} = \tilde{\mathbf{U}}-\mathbf{U}$, which is governed by the linearized Navier-Stokes equations,
\begin{subequations}
    \label{Eqn:LNS}
    \begin{eqnarray}
	    \frac{\partial \mathbf{u}}{\partial t} + \left(\mathbf{U} \cdot \boldsymbol{\nabla} \right) \mathbf{u} + \left(\mathbf{u} \cdot \boldsymbol{\nabla} \right) \mathbf{U} &=& - \boldsymbol{\nabla} p + \frac{1}{Re} \nabla^2 \mathbf{u}, \\
        \boldsymbol{\nabla} \cdot \mathbf{u} &=& \mathbf{0}. 
    \end{eqnarray}
\end{subequations}
The initial condition of the linearized equations is $\mathbf{u}_0 = \tilde{\mathbf{U}}_0 - \mathbf{U}_0$.

Here we focus on the contribution of an instantaneous observation to the estimation of the initial condition, and hence we adopt the instantaneous cost function (\ref{Eqn:CostFunctionNL}).
In this manner, we can evaluate the influence of the observation time $t_m$.  
Extension to the full assimilation window involves integration over time, which is straightforward.
In terms of the new variable $u$, the cost function (\ref{Eqn:CostFunctionNL}) becomes,
 \begin{equation}
 \label{Eqn:CostFunction}
        \mathcal{J}_u(t_m) = \frac{1}{2S}\int_S \left( \frac{\partial u}{\partial y} \right)^2_{(\mathbf{x}_m,t_m)} dx_m dz_m,
 \end{equation}
and similar expressions can be written for $\mathcal{J}_w$ and $\mathcal{J}_p$.
Note that at the true solution $\mathbf{u}_0 = \tilde{\mathbf{U}}_0 - \mathbf{U}_0=\mathbf{0}$, and the gradient of the cost function (generically denoted as $\mathcal{J}$) vanishes,
\begin{equation}
 	\frac{\partial \mathcal{J}}{\partial \tilde{\mathbf{U}}_0}\Big\rvert_{\tilde{\mathbf{U}}_0 =\mathbf{U}_0}  = \frac{\partial \mathcal{J}}{\partial \mathbf{u}_0}
\Big\rvert_{\mathbf{u}_0 = \mathbf{0}} = \mathbf{0}.
\end{equation}
Therefore, the difficulty of the state estimation is directly tied to the second-order derivative, or Hessian matrix, of the cost function.
Note that the Hessian matrix is positive semi-definite because the cost function (\ref{Eqn:CostFunction}) is always non-negative and quadratic in $u$.
If the Hessian is well conditioned, converging from the vicinity of the true solution will be straightforward; 
Conversely, an ill-conditioned Hessian obstructs the optimization algorithm from finding the true state, even if the initial guess is infinitesimally close to the true solution. 

The cost function $\mathcal{J}_u$ depends on $\mathbf{u}$, and involves solving the linearized forward Navier-Stokes equations using the initial condition $\mathbf{u}_0$ and registering observations. 
Computing the Hessian using this forward relation is very expensive, since each possible spatial location in the initial condition and each velocity component must be perturbed independently, and therefore $O(nx \times ny \times nz \times 3)$ simulations are required, each evolved up to the time of the observation $t_m$.
A more efficient approach is sought using the duality relation of the linearized Navier-Stokes equations and their adjoint.  

Consider the Lagrangian identity, 
\begin{equation}
\label{Eqn:duality}
            \left[\mathbf{u}(t = t_m), \boldsymbol{\phi}\left(\mathbf{x}_m\right)\right] = \left[ \mathcal{A}\mathbf{u}_0, \boldsymbol{\phi}\left(\mathbf{x}_m\right) \right] = \left[ \mathbf{u}_0, \mathcal{A}^*\boldsymbol{\phi}\left(\mathbf{x}_m\right) \right] = \left[\mathbf{u}_0, \mathbf{u}^*(\tau = t_m;\mathbf{x}_m,t_m)\right],
\end{equation}
where $\left[\mathbf{a},\mathbf{b}\right] =\displaystyle\int_V \mathbf{a}^T\mathbf{b} dV$ is the spatial inner product, and $\boldsymbol{\phi}$ is the kernel for evaluating observations. 
For example, when observing $\left(\partial U/\partial y\right)_{\text{wall}}$ at one location, the operator $\boldsymbol{\phi}$ is the wall-normal derivative of the streamwise velocity component at the wall at $\mathbf{x}_m$.
The forward linearized operator $\mathcal{A}$ represents solving equations (\ref{Eqn:LNS}) to time $t = t_m$, and $\mathcal{A}^*$  represent solving the adjoint linearized equations,
\begin{subequations}
    \label{Eqn:AdjointLinearized}
    \begin{eqnarray}
    \boldsymbol{\nabla} \cdot \mathbf{u}^* &=& 0,\\
    \frac{\partial \mathbf{u}^*}{\partial \tau} +\left(\boldsymbol{\nabla} \mathbf{U}\right)\cdot \mathbf{u}^* - \left(\mathbf{U} \cdot \boldsymbol{\nabla} \right) \mathbf{u}^* &=& \boldsymbol{\nabla} p^* + \frac{1}{Re} \nabla^2 \mathbf{u}^*,\\
    \mathbf{u}^*(\tau = 0) &=& \boldsymbol{\phi}(\mathbf{x}_m),
\end{eqnarray}
\end{subequations}
where $\tau = t_m-t$ is the reverse time, and the interval $0\le t \le t_m$ is the same time horizon as the forward equations.
It is important to note the difference between the present adjoint equations (\ref{Eqn:AdjointLinearized}) and the earlier (\ref{Eqn:Adjoint_forcing}); here the equations do not feature a forcing term by the cost function and their initial condition is the observation kernel. 
This adjoint field is therefore not the gradient of the cost function. Instead, it is best understood in terms of the Lagrangian identity (\ref{Eqn:duality}) which provides two physical interpretations of deviations in estimated and true observations: 
The first is the inner product of the forward field $\mathbf{u}$ at the observation time with the observation kernel $\boldsymbol{\phi}$; 
The second is the inner product of the error in the initial state $\mathbf{u}_0$ with the adjoint field $\mathbf{u}^*$.
The error in the initial state influences the final observations if and only if $\mathbf{u}_0$ is non-zero within the support of $\mathbf{u}^*$.
In other words, $\mathbf{u}^*$ represent the \emph{domain of dependence} of the observation evaluated at $\left(\mathbf{x}_m, t_m\right)$.

With knowledge of the adjoint field $\mathbf{u}^*$, we can rewrite the cost function with observation kernel $\boldsymbol{\phi}$ at time $t_m$ as,
\begin{equation}
    \label{Eqn:CostFunction_ustar}
    \mathcal{J}(\mathbf{u}_0;t_m) = \frac{1}{2S}\int_S \left[ \mathbf{u}_0, \mathbf{u}^* \right]^2 dx_m dz_m.  
\end{equation}
The quadratic relation between $\mathcal{J}$ and $\mathbf{u}_0$ thus becomes explicit, and facilitates both the evaluation of the gradient and the derivation of the Hessian, 
 \begin{eqnarray}
 \frac{\partial \mathcal{J}}{\partial \mathbf{u}_0} &=& \frac{1}{S} \int_S\left[\mathbf{u}_0, \mathbf{u}^*\right] \mathbf{u}^* dx_m dz_m dt_m, 
 \nonumber \\
 \mathcal{H}(\mathbf x_1, \mathbf x_2;t_m) \equiv \frac{\partial^2 \mathcal{J}}{\partial \mathbf{u}_0\partial \mathbf{u}_0} &=& \frac{1}{S}\int_S \mathbf{u}^*\mathbf{u}^* dx_m dz_m.
 \label{Eqn:Hessian1}
\end{eqnarray}
At optimality, $\mathbf{u}_0 = \mathbf{0}$ and the gradient vanishes while the Hessian is finite and is a measure of the cross-correlation of the adjoint field,  $\mathbf{u}^*\mathbf{u}^*$.
For different choices of the observation kernel $\boldsymbol{\phi}$, be it the wall-normal gradient or sampling of the local pressure, we define  $\mathcal{J}_{u},\mathcal{J}_{w}$, $\mathcal{J}_{p}$ and use the above outlined procedure to evaluate the associated Hessian matrices, $\mathcal{H}_u, \mathcal{H}_w$, $\mathcal{H}_p$.
Written explicitly, the Hessian has the form, 
\begin{equation}
\mathcal{H}_{ij}(\mathbf{x}_1,\mathbf{x}_2; t_m) = \frac{1}{S}\int_S u_i^*(\mathbf{x}_1,\tau = t_m ;\mathbf{x}_m, t_m) u_j^*(\mathbf{x}_2, \tau = t_m;\mathbf{x}_m, t_m) dx_m dz_m.
\end{equation}
Since the adjoint field is the solution to (\ref{Eqn:AdjointLinearized}) starting from the observation kernel, or  $\mathbf{u}^* = \mathcal{A}^* \mathbf{u}^*(\tau = 0) = \mathcal{A}^* \boldsymbol{\phi}(\mathbf{x}_m )$, it is a function of the observation location $\mathbf{x}_m$.  For this reason, the Hessian involves an integral over $dx_m dz_m$.
Performing the integral (\ref{Eqn:Hessian1}) at a specific $t_m$ requires solving the adjoint for every possible ($x_m, z_m$) pair. 
We can however exploit homogeneity in the horizontal $x$ and $z$ directions, and sample a number of $(x_m,z_m)$ locations on the wall. 
We also sampled different starting times of the adjoint simulation, taken as different realizations of the turbulent channel flow.
Periodicity of the adjoint fields in the horizontal plane facilitates expressing the Hessian in Fourier space,
\begin{equation}
    \label{Eqn:HessianFourier}
	\hat{\mathcal{H}}(k_x, k_z ; t_m) =\int_S\hat{\mathbf{ u}}^*(k_x,k_z,\tau = t_m;\mathbf{x}_m, t_m)\hat{\mathbf{u}}^*( k_x,k_z,\tau = t_m;\mathbf{x}_m, t_m) dx_m dz_m.
\end{equation}
Note that this form does not alter the eigenvalues or eigenvectors of the Hessian, and is computationally more efficient because the original convolution integral in (\ref{Eqn:Hessian1}) is now replaced by a product in Fourier space (\ref{Eqn:HessianFourier}). This form also allows us to examine particular wavenumbers.

In total, for every observation time $t_m$, 512 adjoint simulations, or samples, were used to evaluate the ensemble average of the Hessians associated with observing each of $\partial U/\partial y \vert_{wall}$, $\partial W/\partial y\vert_{wall}$ and $P\vert_{wall}$. 
Although we significantly reduce the computational cost for evaluating the Hessian matrix by utilizing the Lagrangian identity, the performed adjoint simulations required approximately 1.2 million CPU hours for each Reynolds number.

\section{Results}
\label{Sec:Results}
This section initially focuses on turbulent channel flows at $Re_{\tau} = 180$ and $590$. 
Ensemble averaged adjoint fields $\overline{\mathbf{u}^*}$ will be reported for different types of wall data and reverse times. 
Appropriate normalization demonstrates the similarity across different Reynolds numbers.
The results also highlight the domains of dependence of different wall sensing modalities and their change with the time separation between the observation and the initial state. 
Eigen-decomposition of the Hessian will be performed, and the eigen-values and vectors at different wave numbers and reverse times will provide a unique perspective on the problem of initial-state estimation.
The section concludes with a discussion of the long-time statistical properties of the adjoint fields for select Reynolds numbers from table \ref{TABLE:Resolution}, $Re_{\tau} = \{180, 590, 1000\}$, and their implications.

\subsection{
Domain of dependence of isolated wall observation
}
\label{Sec:AdjointImpulse}

\begin{figure}
    \centering
    \includegraphics[width = \textwidth]{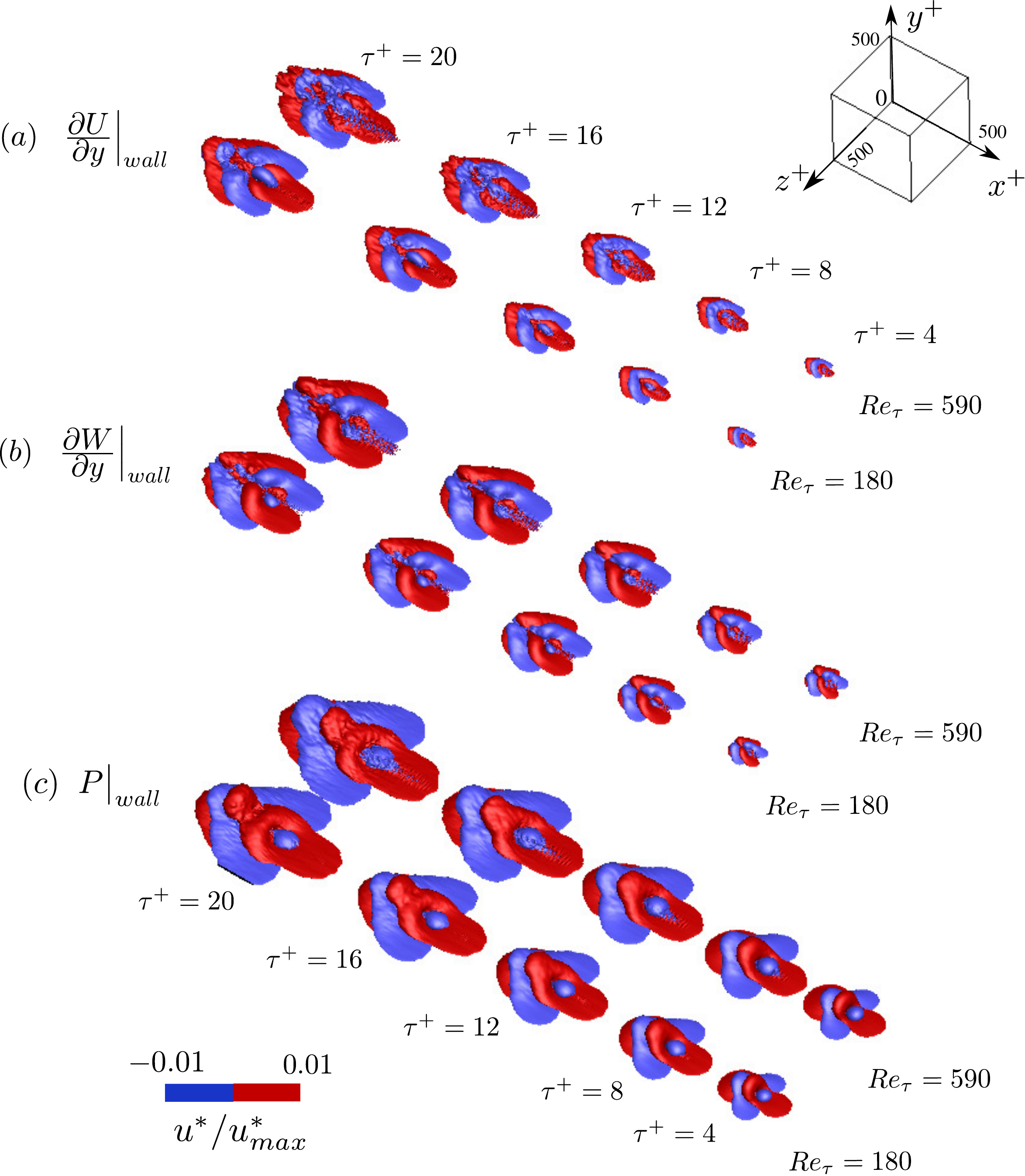}
    \caption{Three dimensional iso-surfaces of the streamwise velocity in the ensemble-averaged adjoint fields for different measurement times $t_m^+$ when observing (a) $\frac{\partial U}{\partial y}\vert_{wall}$, (b) $\frac{\partial W}{\partial y}\vert_{wall}$, and (c) $P\vert_{wall}$, plotted at $\tau^+ = t_m^+$.}
    \label{Fig:isoSurfaces}
\end{figure}

The Lagrangian identity (\ref{Eqn:duality}) provides an interpretation of the adjoint field, where the support of $\mathbf{u}^*(\tau=t_m)$ can be viewed as the domain of dependence for an observation.
We evaluated the adjoint associated with different observation locations $\mathbf{x}_m$ on the wall, each evolving in backward time using the stored forward velocity field (see \ref{Eqn:AdjointLinearized}).  For each observation location, we have also considered different starting times.  
Due to the homogeneity of channel-flow turbulence in the horizontal plane and time, the structures of all these adjoint fields $\mathbf{u}^*(\tau = t_m)$ are similar within a coordinate shift.  
The ensemble average of the adjoint fields was evaluated and is reported in figure \ref{Fig:isoSurfaces}. 

\begin{figure}
    \centering
    \includegraphics[width=0.8\textwidth]{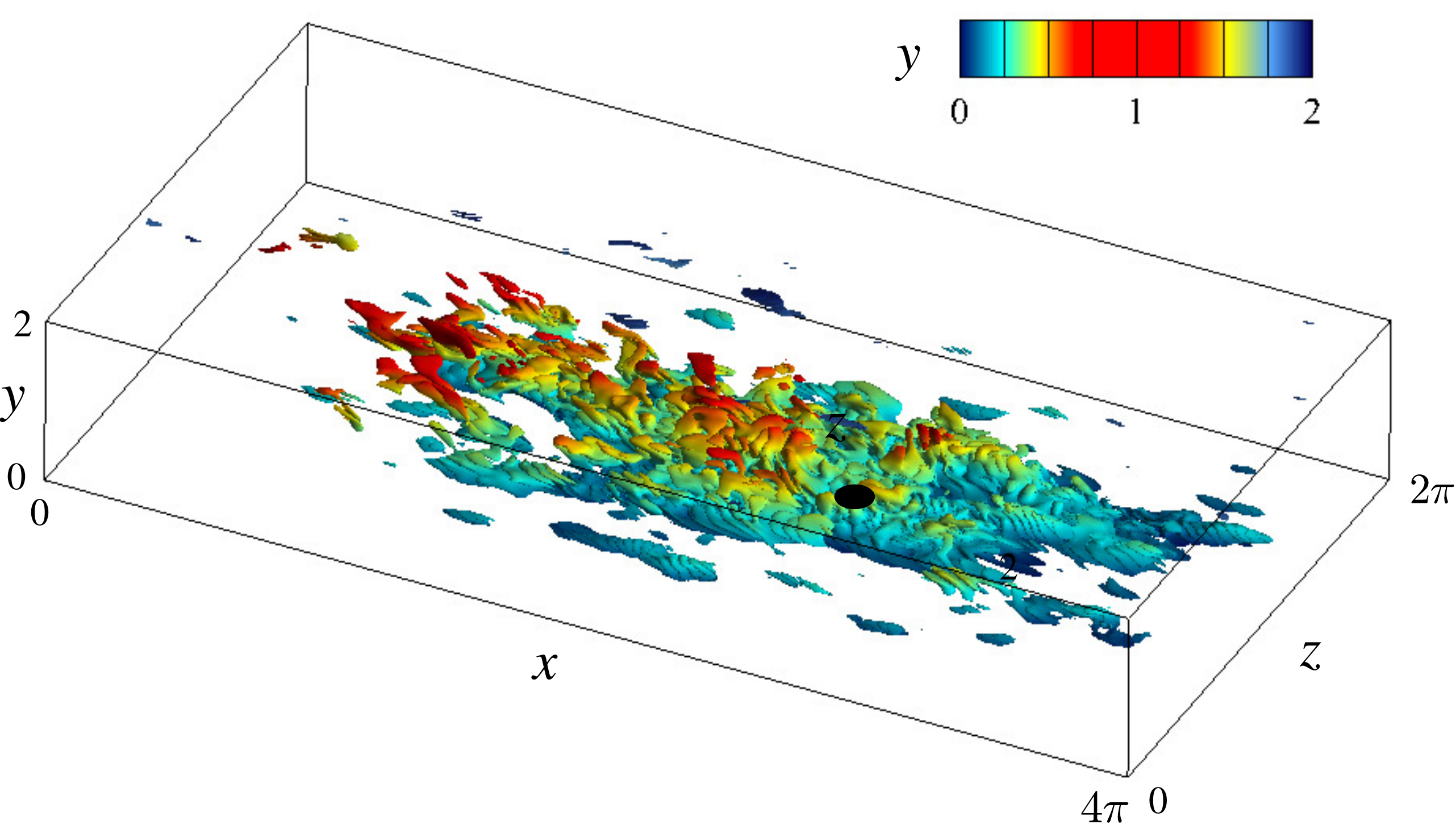}
    \caption{Adjoint field at $\tau^+ = t_m^+ = 220$ due to an impulse of $\frac{\partial u}{\partial y}\vert_{wall}$ at the wall from ($\bullet$) the observation location $(x_m,z_m)=(8\pi/3, \pi)$ for $Re_{\tau} = 180$.
    The iso-surfaces show $u^* = \pm 0.001u_{max}^*$, and are colored by the vertical distance from the wall.}
    \label{Fig:LongTimeAdjoint}
\end{figure}

Three different types of observations are examined, namely $\partial U/\partial y \vert_{wall}$, $\partial W/\partial y\vert_{wall}$ and $P\vert_{wall}$. 
For each measurement modality, 512 samples were included in the ensemble average.
Figure \ref{Fig:isoSurfaces} shows that the fields are similar at the two Reynolds number, when visualized using viscous scaling.  
No matter the choice of wall observations, the adjoint structures all advect upstream of the observation point and expand as a function of reverse time, or for longer separation between the observation and the initial state.
The quantity plotted is the streamwise adjoint velocity, specifically $u^*/u^*_{max}$.  
The spanwise symmetry when observing $\partial U/\partial y \vert_{wall}$ and $P\vert_{wall}$ can be contrasted to the anti-symmetry when measuring $\partial W/\partial y\vert_{wall}$;  in the first two cases, initial disturbances to the forward $U$ velocity at either side of the measurement can not be distinguished, but they have opposite influences on a measurement of the spanwise stress.
In addition, due to the non-local nature of pressure, the adjoint structure associated with wall-pressure observations, $P\vert_{wall}$, are appreciably larger than those associated with the wall stresses, especially for shorter times. 
This observation is congruent with the finding by \cite{bewley2004skin} that including wall-pressure observations in addition to the surface stresses improves the accuracy of state estimation.

Figure \ref{Fig:LongTimeAdjoint} shows the adjoint field for an even longer separation between the observation and the initial state, $\tau^+ = t_m^+ = 220$ at $Re_{\tau} = 180$.  The iso-surfaces of adjoint streamwise velocity form a large chaotic patch that resembles the familiar turbulent spots from transitional flows \citep{cantwell_1978,marxen2019turbulence}, but with the opposite orientation and direction of propagation. 
The reverse-spot spans most of the horizontal plane and it becomes difficult to visually infer the location of its inception, which is the original observation kernel. The long-time behavior of the adjoint field becomes independent of its initial condition, and is only meaningful to examine statistically.  
In addition, due to the chaotic nature of the underlying turbulent field $\mathbf{U}$, the convergence behavior of the adjoint statistics quickly deteriorates and the required number of samples rapidly increases with larger $t_m$ \citep[see e.g.][]{eyink2004ruelle,chandramoorthy2019feasibility}. 
A discussion of the asymptotic behavior of the adjoint field and its impact on the state estimation problem is provided in \S\ref{Sec:AdjointAsymptotic}, after we examine eigen-properties of the Hessian matrix at short times and the implication for state estimation (c.f.\,figure \ref{Fig:SampleReconstruction}).

\subsection{Eigenvalues and eigenvectors of the Hessian}
\label{Sec:EigenAnalysis}

\begin{figure}
    \centering
    \includegraphics[width = 0.8\textwidth]{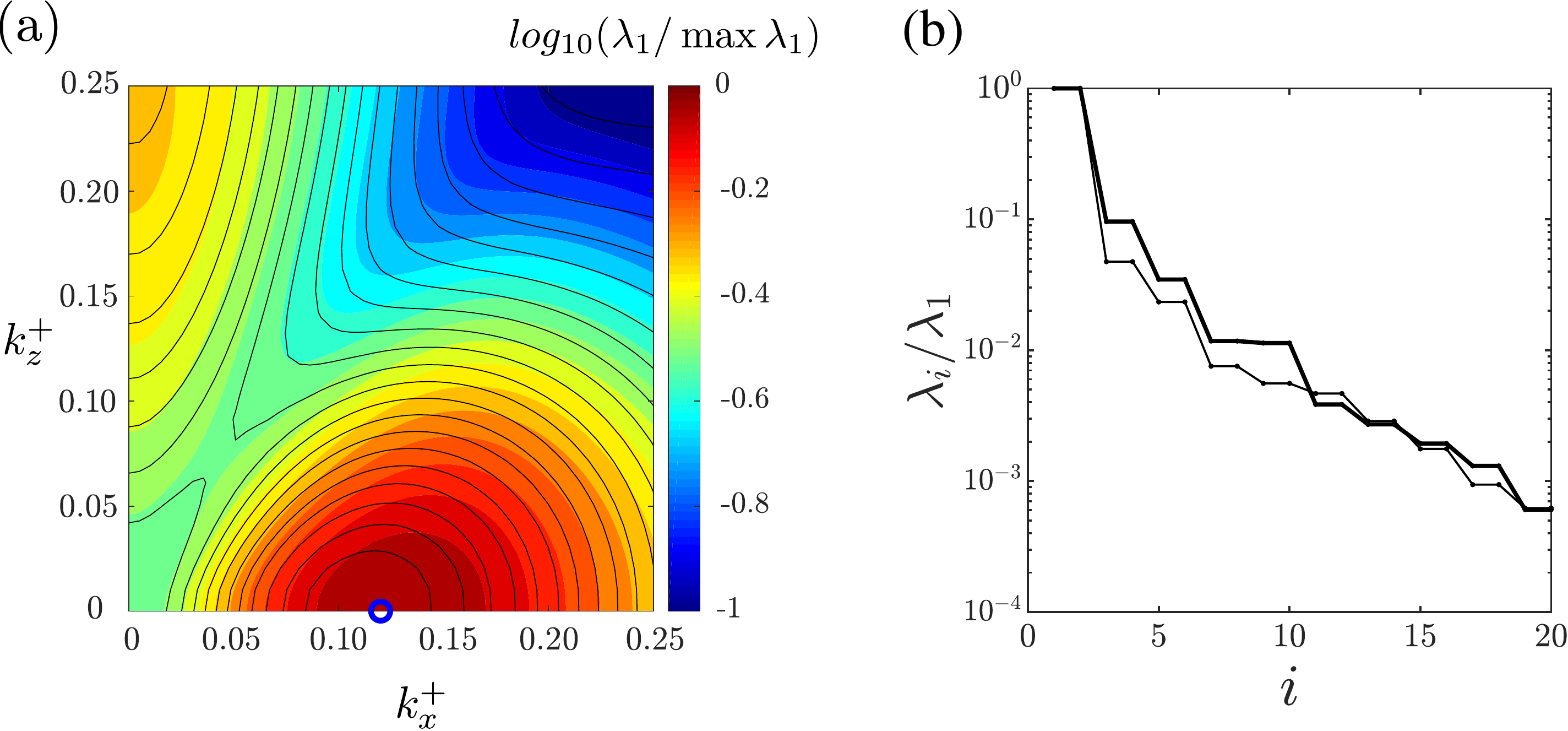}
    \caption{(a) The logarithm of the largest eigenvalue of the Hessian matrix $\hat{\mathcal{H}}_u(k_x^+, k_z^+; t_m^+)$ at $t_m^+ = 4$. Color contours show the eigenvalues for $Re_{\tau} = 180$ and line contours correspond to $Re_{\tau} = 590$. (b) Eigenvalue spectra at the wavenumber pair $(k_x^+,k_z^+) = (0.12, 0)$; thicker line is the higher Reynolds number.}
\label{Fig:Eigen_u_0p2}
\end{figure}

Due to periodicity in $x$ and $z$ directions, the eigenmodes of the Hessian are Fourier modes in these dimensions and their $y$-dependence can be evaluated from $\hat{\mathcal{H}}(k_x^+, k_z^+; t_m^+)$ for different wavenumber pairs $(k_x^+,k_z^+)$.
The results can be unified for $Re_{\tau} = 180$ and $590$ by adopting viscous scaling. 
With reference to the eigen-values and vectors of the Hessian, we can then explain the capacity to reconstruct different wavenumber components of the initial flow state.
The contours in figure \ref{Fig:Eigen_u_0p2} correspond to the largest eigenvalue of $\hat{\mathcal{H}}_u(k_x^+,k_z^+; t_m^+)$, the Hessian for instantaneous observation of $\partial U/\partial y \vert_{wall}$, as a function of the wavenumber vector. A short observation time is considered, $t_m^+ = 4$.  
The two Reynolds numbers $Re_{\tau} = 180$ and $590$ are compared using the color and line contours, respectively, which demonstrates their agreement when scaled using viscous units.
The maximum eigenvalue generally decays with increasing magnitude of the horizontal wavenumber vector $(k_x^+, k_z^+)$, and the supremum corresponds to a two-dimensional mode $(k_x^+, k_z^+) = (0.12, 0)$.  
Figure \ref{Fig:Eigen_u_0p2}$b$ shows the entire eigenvalue spectra of $\hat{\mathcal{H}}_u(k_x^+,k_z^+;t_m^+)$ at that wavenumber pair, for both Reynolds numbers. 
The eigenvalues appear in pairs, and decay exponentially, which demonstrates that the sensitivity of the system is dominated by few directions associated with the leading eigenvectors: a typical feature of ill-conditioned systems.

\begin{figure}
    \centering
    \includegraphics[width = 0.8\textwidth]{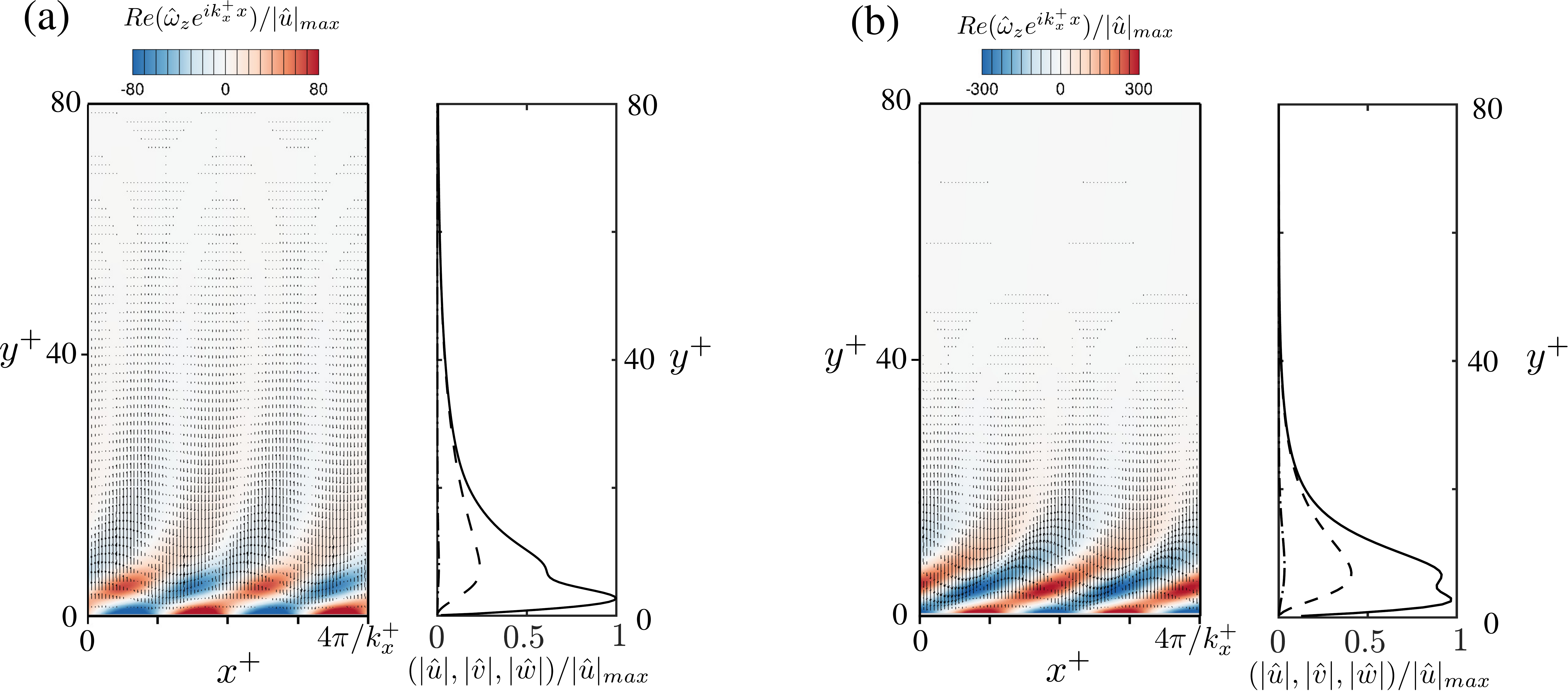}
    \caption{Visualizations of the leading eigenmodes of the Hessian matrix with wavenumber $(k_x^+,k_z^+)=(0.12, 0)$ for (a) $Re_{\tau} = 180$ and (b) $Re_{\tau} = 590$. 
    Side view contours of the spanwise vorticity $\omega_z \equiv \frac{\partial {v}}{\partial x} - \frac{\partial {u}}{\partial y}$ and arrows of $(u,v)$ components are shown.
    Line plots show the mode shapes as a function of $y$ for $\hat{u}$ (\sampleline{}), $\hat{v}$ (\sampleline{dashed}) and $\hat{w}$ (\sampleline{dash pattern=on .7em off .2em on .2em off .2em}) components normalized by $|\hat{u}|_{max}$.
    }
    \label{Fig:ModeShape}
\end{figure}

The profiles of select eigenvectors, $\left(\hat{u}(y), \hat{v}(y),\hat{w}(y)\right)$, are examined in figure \ref{Fig:ModeShape}.  
The mode associated with the supremum eigenvalue at $(k_x^+, k_z^+) = (0.12,0)$ corresponds to spanwise homogeneous rolls. 
The roll peaks at $y^+ \approx 3$ and generates counter vorticity at the wall that most effectively impacts the wall observation of streamwise shear stress.
This eigenmode should not be interpreted using the conventional wisdom from forward simulations of turbulence and knowledge of its structures; the interpretation should be based on the notion of the adjoint.  
The eigenfunction shows the direction along which the cost function has the largest curvature, and hence the measurement is most sensitive to a perturbation in this direction at $\tau = t_m$ earlier in time.  
When this time is short, as is the case in figure \ref{Fig:ModeShape}, the forward dynamics do not have sufficient time to act on the evolution of the initial disturbance in a manner that can lead to a significant sensor signal.  
As such, the eigenfunction is an unfamiliar initial disturbance that would maximize the sensor response, but may not necessarily be common in simulations of turbulence.  
It is nonetheless the leading direction in the process of minimizing the cost function during an adjoint-variational data assimilation using measurement of streamwise shear stress at $t_m^+ = 4$.

\begin{figure}
    \centering
    \includegraphics[width = 0.9\textwidth]{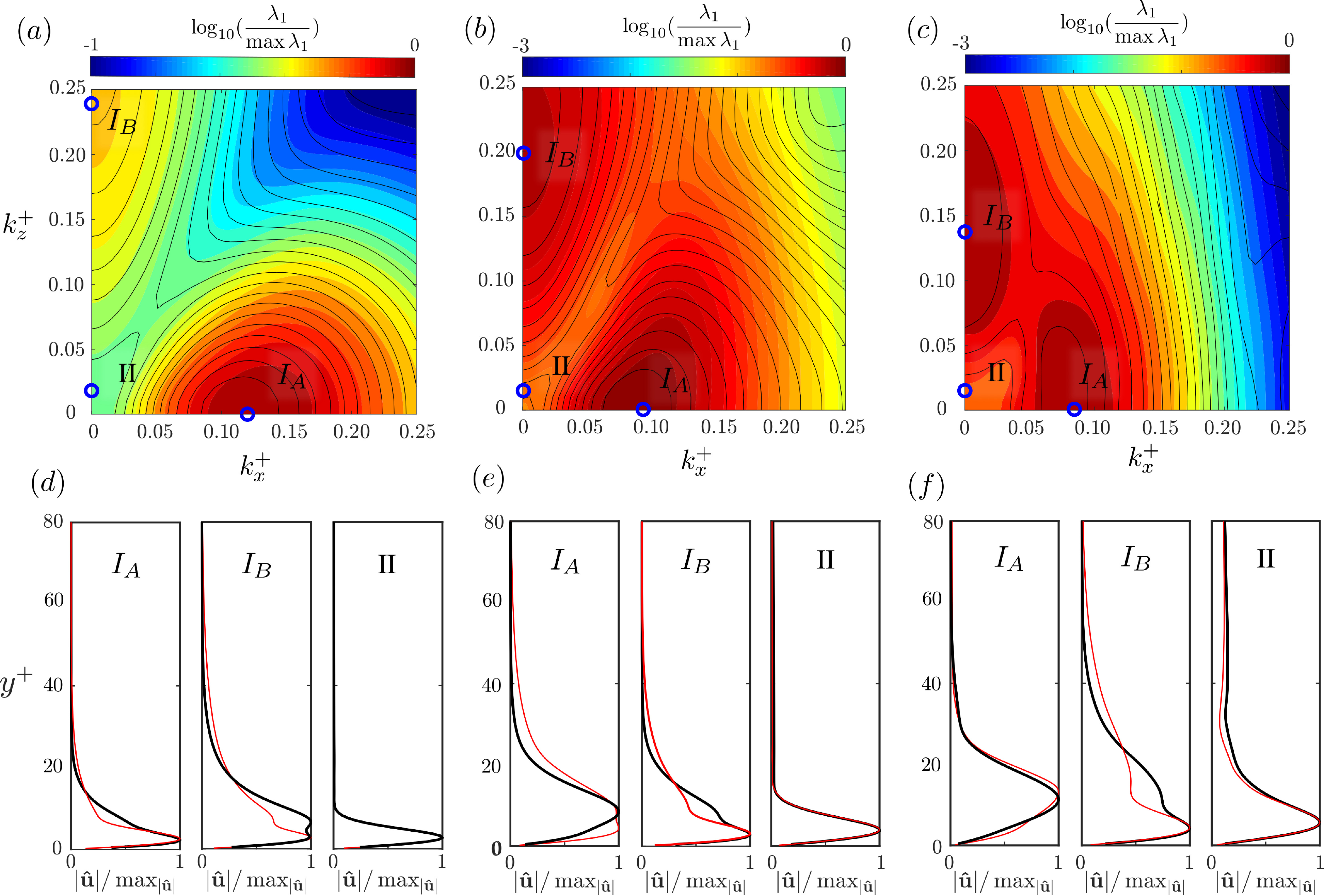}
    \caption{The logarithm of the largest eigenvalue of the Hessian matrix for observing $\partial U/\partial y \vert_{wall}$ instantaneously at (a-c) $t_m^+ = \{4, 8, 20\}$. 
    Color and line contours are for $Re_{\tau} = 180$ and $590$, respectively.
    The eigenvalues are reported as a function of the horizontal wavenumber vector $(k_x^+, k_z^+)$ and are normalized by the respective supremum. 
    (d-f) Profiles of the eigenfunction $|\hat{\mathbf{u}}|$ associated with the wavenumbers marked in panels (a-c). 
    Thin red lines correspond to $Re_{\tau} = 180$ and thick black lines are for $Re_{\tau} = 590$.
    }
    \label{Fig:ElevationMap2}
\end{figure}

Figure \ref{Fig:ElevationMap2}(a,b,c) show contours of the normalized eigenvalues of the Hessian as a function of $(k_x^+,k_z^+)$, for increasing observation time $t_m^+$. 
Here the Hessian matrices are associated with instantaneous observations of $\partial U/\partial y \vert_{wall}$ at $t_m^+$. 
In each panel, we identify three important wavenumber pairs, and plot the associated eigenmodes in panels (d,e,f).
Modes $I_A$ and $I_B$ correspond to the two local peaks of the eigenvalues on the horizontal and vertical axis, which represent spanwise and streamwise rolls , respectively.
Mode $II$ represents much large structures with $\lambda_z^+  =\mathcal{O}(300)$. 
Compared to other eigenfunctions, mode $II$ demonstrates finite sensitivity of wall measurement to the flow beyond the buffer layer towards the core of the channel when the measurement time $t_m^+$ is large (panel \ref{Fig:ElevationMap2}(f)).

Five remarks are notable in connection with figure \ref{Fig:ElevationMap2}: 
(i) The supremum of the eigenvalues for observing $\partial U/\partial y \vert_{wall}$ remains two dimensional for all times;
(ii) At short time, the most efficient perturbations to influence the wall measurement have large wavenumbers and the corresponding eigenvectors are clustered near the wall, representing an immediate sensitivity that is not significantly influenced by the underlying dynamics of the flow. 
(iii) However, at longer times, such high wavenumber structures would be dampened due to viscous effects and therefore the most prominent eigenvalues shifts toward longer streamwise and spanwise wavelengths.
These larger structures survive longer times, and can be amplified by the flow dynamics over the time duration between initial condition and measurement.  
(iv) While all modes expand farther from the wall at longer observation times, most eigenfunctions are vanishing above the buffer layer\textemdash a detail that we quantify in appendix \ref{Sec:Elevation} (see figure \ref{Fig:yloc}).
The important implication is that wall observations of the streamwise stress are not sensitive to many of the turbulence scales above the buffer layer, especially large values of $(k_x^+, k_z^+)$.  
(v) A very important exception is in a small region near mode II where the eigenfunctions maintain a finite value in the core of the channel; these modes correspond to the sensitivity of observations of the streamwise wall stress to outer large-scale structures.  These elongated motions are immune to the sheltering effect of the strong near-wall shear \citep[see][for discussion of shear sheltering]{hunt1999perturbed,zaki2009shear}, and their influence on the wall stress is consistent with earlier efforts focused on the forward evolution of turbulence \citep{hwang2016inner, mathis2009large, abe2004very, you2019tbl}.
The implication of last two observations (iv and v) is evident in figure \ref{Fig:SampleReconstruction}, where only the large-scale motions in the logarithmic layer are accurately reconstructed and the finer scales of turbulence are absent from the estimated state.  

\begin{figure}
    \centering
    \includegraphics[width = \textwidth]{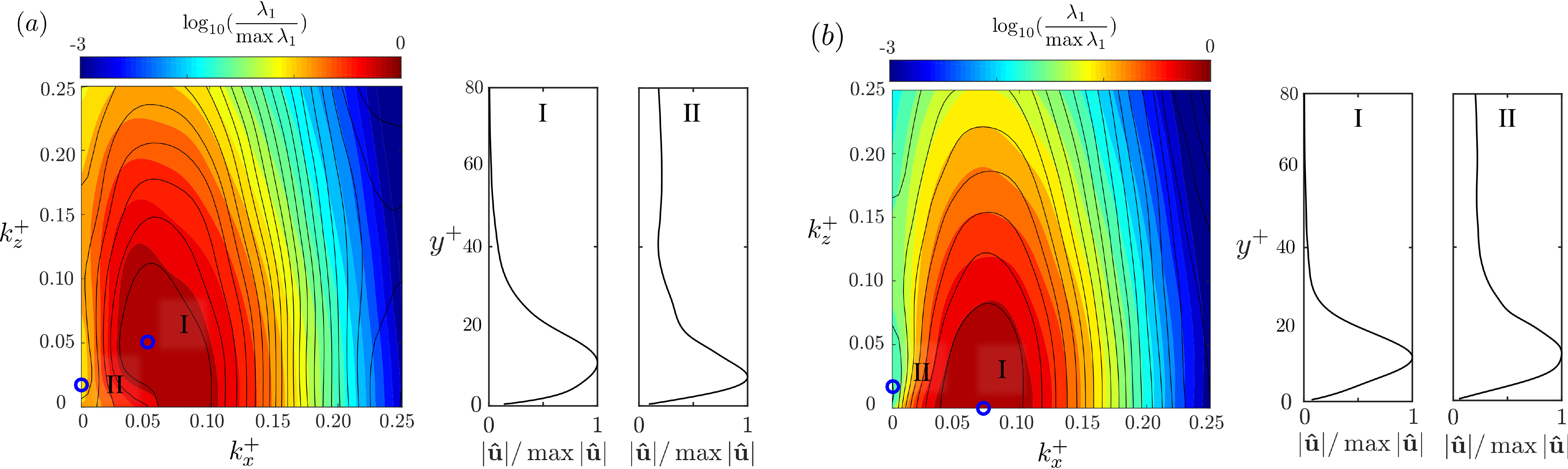}
    \caption{Contours of the logarithm of the largest eigenvalue of the Hessian matrix for observing (a) $\partial W/\partial y \vert_{wall}$ and (b) $P\vert_{wall}$, instantaneously at $t_m^+ = 20$, as a function of the horizontal wavenumber vector and normalized by the respective suprema. 
    Color and line contours correspond to $Re_{\tau} = 180$ and $590$, respectively.
    Line plots are profiles of the eigenfunctions $|\hat{\mathbf{u}}|$ associated with the marked eigenvalues $I$ and $II$, at $Re_{tau} = 590$.
    }
    \label{Fig:EigenValues}
\end{figure}

The above discussion of the behaviour of the Hessian eigenvalues at long times remains qualitative unchanged when considering wall observations of instantaneous spanwise shear stress or pressure. The key distinctions are made by aid of figure \ref{Fig:EigenValues}, where the associated largest eigenvalues are plotted as a function of $(k_x^+, k_z^+)$, at observation time $t_m^+ = 20$.
The most effective flow structure for observing $\partial W/\partial y \vert_{wall}$ is three-dimensional with wave number $(k_x^+, k_z^+) \approx (0.05, 0.05)$.
As for observing the wall pressure, the eigenvalue spectrum shows sensitivity to spanwise rolls and much weaker sensitivity to streamwise ones, the latter being known to develop without an associated strong pressure perturbation \citep{Phillips1969}. 
The profiles of modes I and II shown in figure \ref{Fig:EigenValues} are similar to those of modes I$_A$ and II in figure \ref{Fig:ElevationMap2}(f) for observing $\partial U/\partial y \vert_{wall}$.
Most importantly, in all three types of observations mode II maintains a finite value beyond the buffer layer, which represents the sensitivity of wall data to the outer large-scale motions.

Mathematically, the eigen-modes of the Hessian matrix can rigorously be interpreted as (a) the leading search directions in adjoint state estimation problem or (b) as the perturbations that wield the largest influence on wall observations.  Such interpretations, while accurate, do not guarantee that such eigen-structures are present, or observable, in developed wall turbulence.
In addition, we argued on physical grounds that short-time observations may not be sensitive to the dynamics that lead to the generation of energetic flow structures in wall turbulence because such structures develop on longer timescale; We also argued on physical grounds that longer-time observations are sensitive to initial disturbances that amplify by the flow dynamics thus having a large wall signature.  
In order to support our interpretation, we can tailor the Hessian analysis to focus on the most energetic modes of developed channel flow at different $(k_x^+, k_z^+)$.  
Specifically, we evaluate the sensitivity of wall observations to 
flow structures $\mathbf {v}$ obtained from a proper orthogonal decomposition (POD) of channel-flow turbulence \citep{Lumley1967, moin_moser_1989,Taira2017modal}.

Assume that the initial perturbation $\mathbf u_0$ is aligned with a POD mode $\mathbf u_0 = \alpha \mathbf {v}$, where $\alpha$ is the amplitude. 
The cost function (\ref{Eqn:CostFunction_ustar}) can be written as,
\begin{equation}
    \label{Eqn:CostFunction_POD}
    \mathcal{J}(\alpha \mathbf {v};t_m) = \frac{1}{2S}\int_S \left[ \alpha \mathbf {v}, \mathbf{u}^* \right]^2 dx_m dz_m = \frac{1}{2S}\int_S \alpha^2 \left[ \mathbf {v}, \mathbf{u}^* \right]^2 dx_m dz_m.  
\end{equation}
Since the POD mode generally satisfies $[\mathbf{ v},\mathbf { v}] = 1$, the inner product $[\mathbf v,\mathbf u^*]$ is the scalar projection of the adjoint field onto the POD mode.
Given $\mathbf{v}$, the cost function only depends on $\alpha$, and the corresponding projected Hessian is the second-order derivative with respect to $\alpha$,
\begin{equation}
    \mathcal{H}_{POD}(t_m) = \frac{\partial^2 \mathcal{J}}{\partial \alpha \partial \alpha} = \frac{1}{S} \int_S \left[ \mathbf {v}, \mathbf{u}^* \right]^2 d x_m d z_m.
\end{equation}
Note that the projected Hessian becomes a scalar, which is equivalent to its eigenvalue and quantifies the sensitivity of wall observations to the POD mode. 
For projection onto POD modes at different $(k_x^+,k_z^+)$, the corresponding Hessian is denoted as ${\mathcal H}_{POD}(k_x^+,k_z^+;t_m)$.

\begin{figure}
    \centering
    \includegraphics[width = 0.6\textwidth]{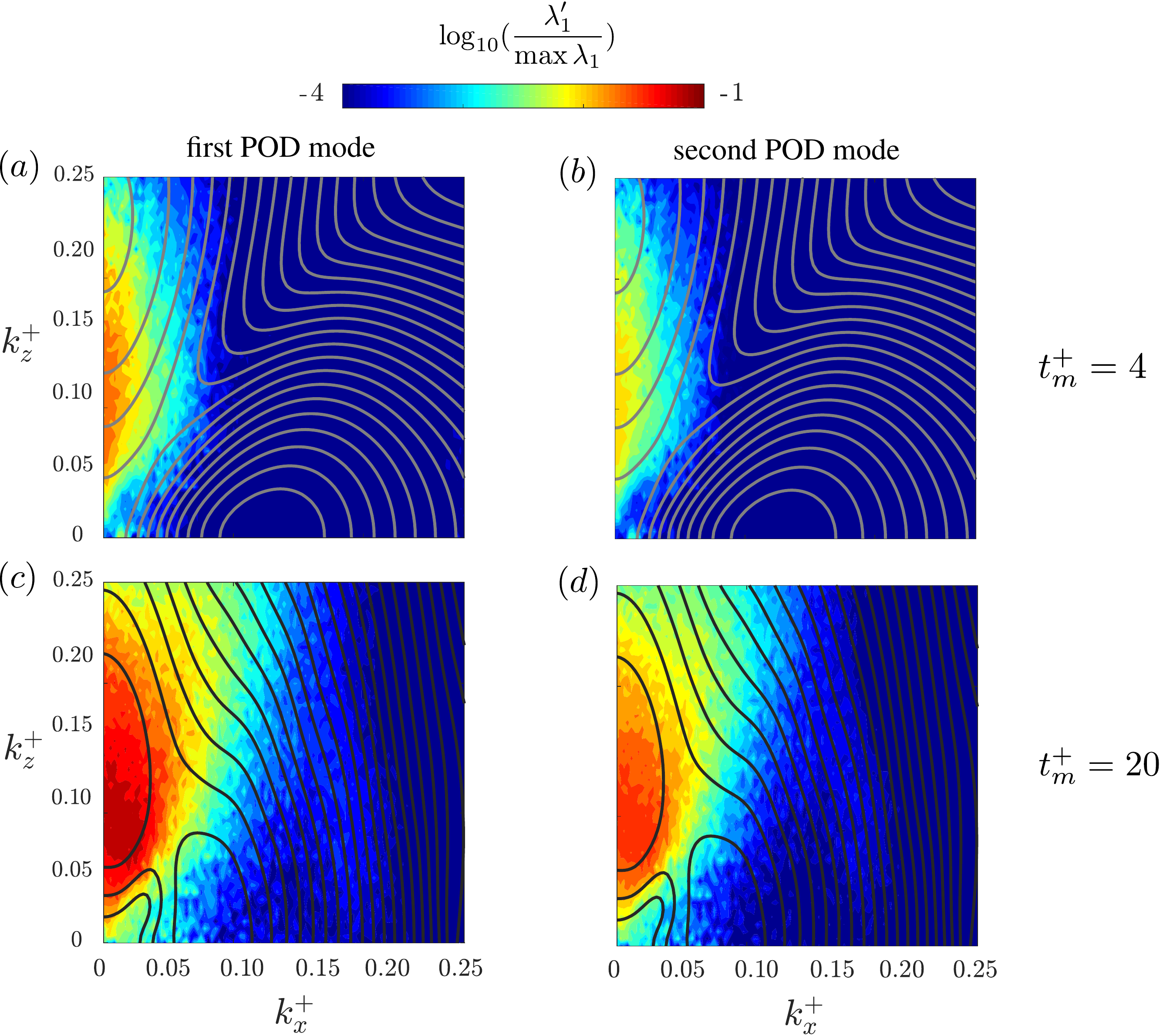}
    \caption{(Color) Eigenvalues of the projected Hessian showing the sensitivity of wall observations of $\frac{\partial U}{\partial y}$ to (left) the first and (right) the second POD modes of turbulent channel flow at $Re_{\tau} = 180$.  Lines are reproduced from figure \ref{Fig:ElevationMap2}, for the original Hessian without projection.  Observation times (top) $t_m^+=4$ and (bottom) $t_m^+=20$. Contour lines show the results without POD projection.}
    \label{Fig:POD}
\end{figure}

Results are reported in figure \ref{Fig:POD} for $Re_{\tau} =  180$.  
The color contours are the eigenvalues $\lambda^{\prime}$ of the Hessian as a function of $(k_x^+,k_z^+)$, when the projection is onto (left) the first and (right) the second POD modes; for comparison, the eigenvalues of the original Hessian, without projection, are included as lines.  The maximum eigenvalues of the latter are used for normalization.  
For early observations, $t_m^+=4$, the spectra of the original and projected Hessian are dissimilar: 
Specifically, the original peak eigenvalue at large $k_x^+$ does not persist, which indicates that the wall observations are not sensitive to the energetic turbulent structures in that wavenumber range. Instead, the sensitivity at large $k_x^+$ is to near-wall small-scale structures, which are the first directions that adjoint-variational data assimilation will attempt to reconstruct. 
For late observations ($t_m^+=20$), the spectra of both the original and projected Hessian are more similar, and are aligned near the peak which is associated with streamwise-elongated structures. 
The implication is that the maximum sensitivity of wall observations is aligned with the most energetic POD modes.  
Adjoint-variational data assimilation using observations at $t_m^+=20$ will therefore target a reconstruction of these energetic structures in the initial state, because the underlying flow dynamics amplify these structures in forward time and thus they lead to the largest impact on the observations.  

The above discussion was framed in terms of the eigenspectrum of the Hessian, and the reported eigenvalues were normalized in all the figures by their supremum at the corresponding observation time.  The time dependence of the supremum is reported in the appendix (figure \ref{Fig:MaxLambda}), and shows long-time exponential amplification due to the Lyapunov behaviour of the \emph{adjoint} system.  As a result, when observations are accumulated during a long time horizon, for example $t_m^+ = 50$ as in section \ref{Sec:AdjointOpt}, the estimation of the initial state is most affected by the late observations.  It is instructive to compare the notions of Lyapunov divergence in the adjoint and forward Navier-Stokes equations: 
Just as in the forward problem where an initial perturbation to the state grows exponentially in time with the Lyapunov exponent, an initial perturbation in the adjoint equations amplifies exponentially in backward time;  For example, at $Re_\tau = 180$ \citet{nikitin2018characteristics} reported a Lyapunov exponent $\sigma^+=0.021$ for the forward problem; the adjoint field has the same exponential rate as we will report in \S\ref{Sec:AdjointAsymptotic}.
The interpretation of the adjoint behaviour is, however, different from the familiar forward problem.  
In the context of the Hessian at optimality, the sensitivity of an observation, which is represented by the adjoint field starting from the observation kernel, leads to exponentially ``larger Hessian" at $\tau = t_m$. 
Not only does the supremum eigenvalue increase, but so does its separation from smaller ones as well (c.f.\,figure \ref{Fig:ElevationMap2}).
As a result, the condition number of the Hessian becomes larger and it becomes much more difficult to solve the state-estimation problem, accurately.  
The important implication is that errors in observations, in particular at later times, can strongly obscure the reconstruction of the initial state.

\subsection{Asymptotic behaviour of the adjoint field \\  \& of the gradient of the cost function}
\label{Sec:AdjointAsymptotic}

The analysis in the previous section focused on evaluation of the Hessian when the gradient of the cost function vanishes.  The results highlighted the difficulty of accurately predicting the full turbulent state, especially beyond the buffer layer, from wall observations. Even when the initial estimate of the flow is near the true state, the wall stress has a diminishing sensitivity to the state with distance from the wall and, in the outer region, can improve the estimation of the large-scale motions only.  
Generally, however, an initial guess of the state may just be the mean-flow profile that is far from the true field. In addition, the assimilation window of observations is often long.
As such, the exponential amplification of the adjoint in backward time can lead to a very large gradient of the cost function with respect to the initial condition, as well as an ill-conditioned Hessian.
These realities impose severe restrictions on the step size in the gradient descent method.
In this section, we examine the kinetic energy of the adjoint field because it is directly proportional to the magnitude of the gradient of the cost function with respect to the initial condition.  
The results will highlight regions of the flow that contribute most to this amplification, and will be contrasted to the evolution of perturbations in the forward field for which we have established understanding.

In \S\ref{Sec:AdjointImpulse}, we examined the adjoint fields due to impulses at the wall (figures \ref{Fig:isoSurfaces} and \ref{Fig:LongTimeAdjoint}), at short observation times.
At long reverse times, the initially localized patches spread, become chaotic and fill the domain.  
Ultimately they reach a statistical state that is independent of the observation type and location.
Therefore, the long-time asymptotic behaviours of the adjoint fields from equations (\ref{Eqn:Adjoint_forcing}) and (\ref{Eqn:AdjointLinearized}) are the same; the former provides the gradient of the cost function and the latter is associated with Hessian matrix.
The asymptotic statistical state can be reached efficiently by performing adjoint computations starting from broadband adjoint velocities throughout the computational domain, which is the approach adopted here.  
It should be noted that the adjoint equations are linear, and hence the initial field is anticipated to grow exponentially in backward time, indefinitely.
This behaviour can be likened to the amplification of perturbations to a forward trajectory: 
While the nonlinear evolution of the perturbations leads to a statistically saturated state, when the linearized Navier-Stokes equations are adopted the perturbations amplify exponentially, unabated.

\begin{figure}
    \centering
    \includegraphics[width = 0.9\textwidth]{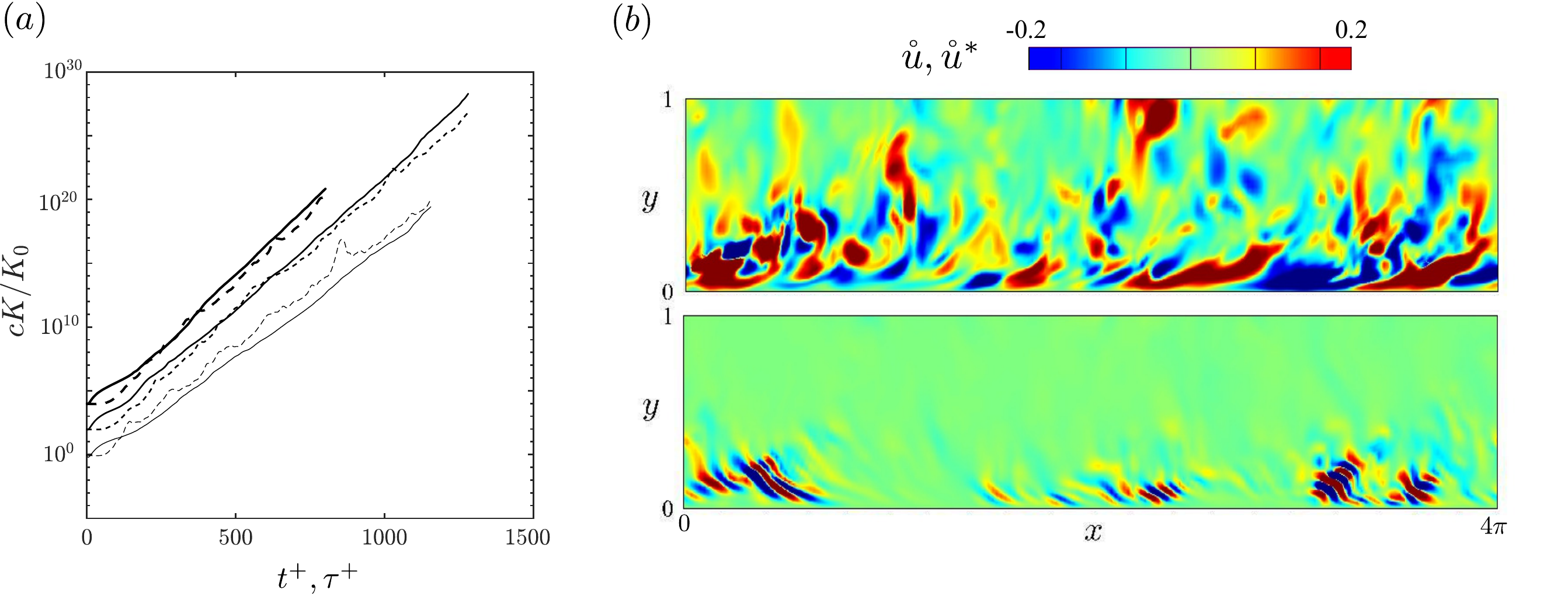}
    \caption{(a) Exponential energy amplification for the (solid) linearized forward and (dashed) adjoint fields. Increasing line thickness and vertical shift correspond to $Re_{\tau} = \{180, 590, 1000\}$.  For each Reynolds number, the Lyapunov exponents are similar for the forward ($\sigma^+ = \{2.11, 2.48, 2.73\}\times 10^{-2}$) and adjoint ($\sigma^+ = \{2.04, 2.33, 2.61\}\times 10^{-2}$). (b) Side views of sample (top) linearized forward and (b) adjoint fields during the statistically stationary state for $Re_{\tau} = 180$.
    }
    \label{Fig:Energy}
\end{figure}

Starting from initial random perturbations that were projected onto solenoidal fields, we simulated the flow evolution using both the linearized forward (\ref{Eqn:LNS}) and adjoint (\ref{Eqn:AdjointLinearized}) Navier-Stokes equations in order to contrast the long-time statistical properties of both models.  Recall that in both cases the base state in the governing equations is the three-dimensional, time-dependent, fully turbulent flow at the corresponding Reynolds number.  
The left panel of figure \ref{Fig:Energy} shows the evolution of the perturbation energy,
\begin{equation}
    K(t)   = \frac{1}{2V}\int_V | \mathbf{u} |^2 dV, \quad\text{and}\quad
    K^*(t) = \frac{1}{2V}\int_V | \mathbf{u}^* |^2 dV. 
\label{Eqn:IntEnergyVol}
\end{equation}
The integration is performed over the full three-dimensional domain, and the exponential growth rate $\sigma^+ \approx 2.11 \times 10^{-2}$ at $Re_{\tau} = 180$ agrees with the value reported by \citet{nikitin2018characteristics} for forward evolution.

\begin{figure}
    \centering
    \includegraphics[width = 0.8\textwidth]{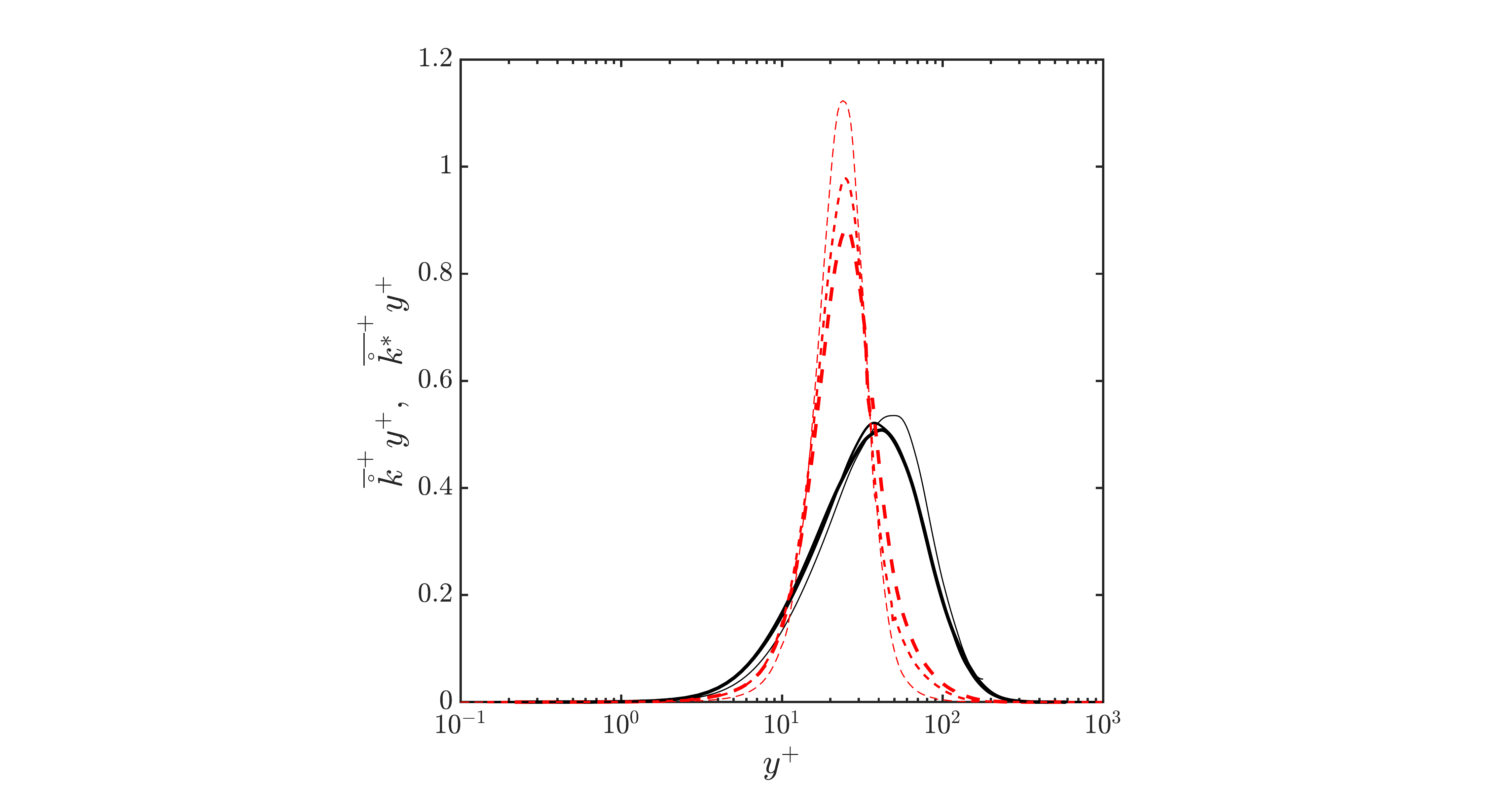}
    \caption{Profiles of the horizontally and time-averaged kinetic energy in the (solid) linearized forward and (dashed) adjoint velocity fields.
    Increasing line thickness corresponds to Reynolds numbers $Re_{\tau} = \{180, 590, 1000\}$.
    Each curve is normalized to have unit integral.}
    \label{Fig:ForwardVsAdjoint}
\end{figure}

Beyond a short-lived transient, the statistical behaviour of the perturbation field within the exponential regime is unchanged, to within a re-scaling. We therefore normalize quantities by their Lyapunov amplification,
\begin{equation}
    \mathring{\mathbf{u}} = \mathbf{u} e^{-\sigma t}, \quad
    \mathring{\mathbf{u}}^* = \mathbf{u}^* e^{-\sigma \tau}, 
\end{equation}
and analyze $\mathring{\mathbf{u}}$ and $\mathring{\mathbf{u}}^*$ as statistically stationary fields.
Figure \ref{Fig:Energy}$b$ shows instantaneous side views of the normalized forward and adjoint perturbation fields for $Re_{\tau} = 180$.
The adjoint field reflects the more localized near-wall behaviour, which is manifest in the form of concentrated high-intensity patches of turbulence.
The horizontally and time-averaged kinetic energy of the normalized variables,
\begin{equation}
    \overline{\mathring{k}^{(*)}}= \frac 12 ~\overline{| \mathring{\mathbf{u}}^{(*)} |^2},
\label{Eqn:IntEnergyAreaTime}
\end{equation}
was evaluated and the wall-normal profiles are plotted in figure \ref{Fig:ForwardVsAdjoint}.   
Each profile is normalized to unit integral, and is plotted multiplied by $y^+$ to ensure that the area under the curve is representative of the integral.
The energy of the forward perturbations peaks in the buffer layer, and decays within the logarithmic region towards the channel center.  
The energy of the adjoint field exhibits narrower support, with a more concentrated peak and much faster decay within the buffer layer. 
The large gradient of the cost function in the state estimation problem is therefore most concentrated in the buffer layer; In addition the Hessian is poorly conditioned because of this region where any observation errors are amplified exponentially due to the chaotic nature of the flow and hence of the adjoint system.

The evolution equations of the kinetic energies of the perturbations are derived by performing the dot product of the linearized forward Navier-Stokes equations (\ref{Eqn:LNS}) with $\mathbf{u}$ and similarly the dot product of the adjoint Navier-Stokes equations (\ref{Eqn:AdjointLinearized}) with $\mathbf{u}^*$.  The resulting equations are, 
\begin{equation}
\begin{aligned}
    \frac{\partial k}{\partial t} = -u_i u_j S_{ij} - \frac{\partial }{\partial x_j} \left(\frac{1}{2}u_i u_i U_j + u_j p -\frac{2}{Re}  u_i s_{ij}\right) - \frac{2}{Re} s_{ij} s_{ij},\\
    \frac{\partial k^*}{\partial \tau} = -u^*_i u^*_j S_{ij} - \frac{\partial }{\partial x_j} \left(- \frac{1}{2}u^*_i u^*_i U_j - u^*_j p^* -\frac{2}{Re}  u^*_i s^*_{ij}\right) - \frac{2}{Re} s^*_{ij} s^*_{ij},
\end{aligned}
\end{equation}
where $k = \frac{1}{2} u_i u_i$ is the kinetic energy, and $s_{ij} = \frac{1}{2} \left(\partial u_i/\partial x_j + \partial u_j / \partial x_i\right)$ and $S_{ij} = \frac{1}{2} \left(\partial U_i / \partial x_j + \partial U_j/\partial x_i\right)$ are the rate of strain tensors for the perturbation and the reference fields, respectively; stars denote adjoint quantities.
We multiple the forward equation by $\exp\left(-2 \sigma t \right)$ and the adjoint by $\exp\left(-2 \sigma \tau \right)$, and average each of them in time and in the homogeneous $x$ and $z$ directions, which yields
\begin{equation}
    2\sigma \bar{\mathring{k}}  =  \underbrace{- \overline{\mathring{u}_i \mathring{u}_j S_{ij}}}_{\mathcal{P}} \underbrace{- \frac{d}{dy}\left(\frac 12 \overline{\mathring{u}_i \mathring{u}_i U_2}  +\overline{\mathring{u}_2 \mathring{p}} - \frac{2}{Re} \overline{\mathring{u}_i \mathring{s}_{i2}}\right)}_{\mathcal{T}} - \underbrace{\frac{2}{Re} \overline{\mathring{s}_{ij}\mathring{s}_{ij}}}_{\epsilon},
\label{Eq:tke_forward}
\end{equation}
\begin{equation}
   2\sigma \overline{\mathring{k}^*} =  \underbrace{- \overline{\mathring{u}^*_i \mathring{u}^*_j S_{ij}}}_{\mathcal{P}^*} \underbrace{- \frac{d}{dy}\left(-\frac 12 \overline{\mathring{u}^*_i \mathring{u}^*_i U_2}  - \overline{\mathring{u}^*_2 \mathring{p}^*} - \frac{2}{Re} \overline{\mathring{u}^*_i \mathring{s}^*_{i2}}\right)}_{
   \mathcal{T}^*} - \underbrace{\frac{2}{Re} \overline{\mathring{s}^*_{ij}\mathring{s}^*_{ij}}}_{\epsilon^*}.
\label{Eq:tke_adjoint}
\end{equation}

\begin{figure}
    \centering
    \includegraphics[width = 0.8\textwidth]{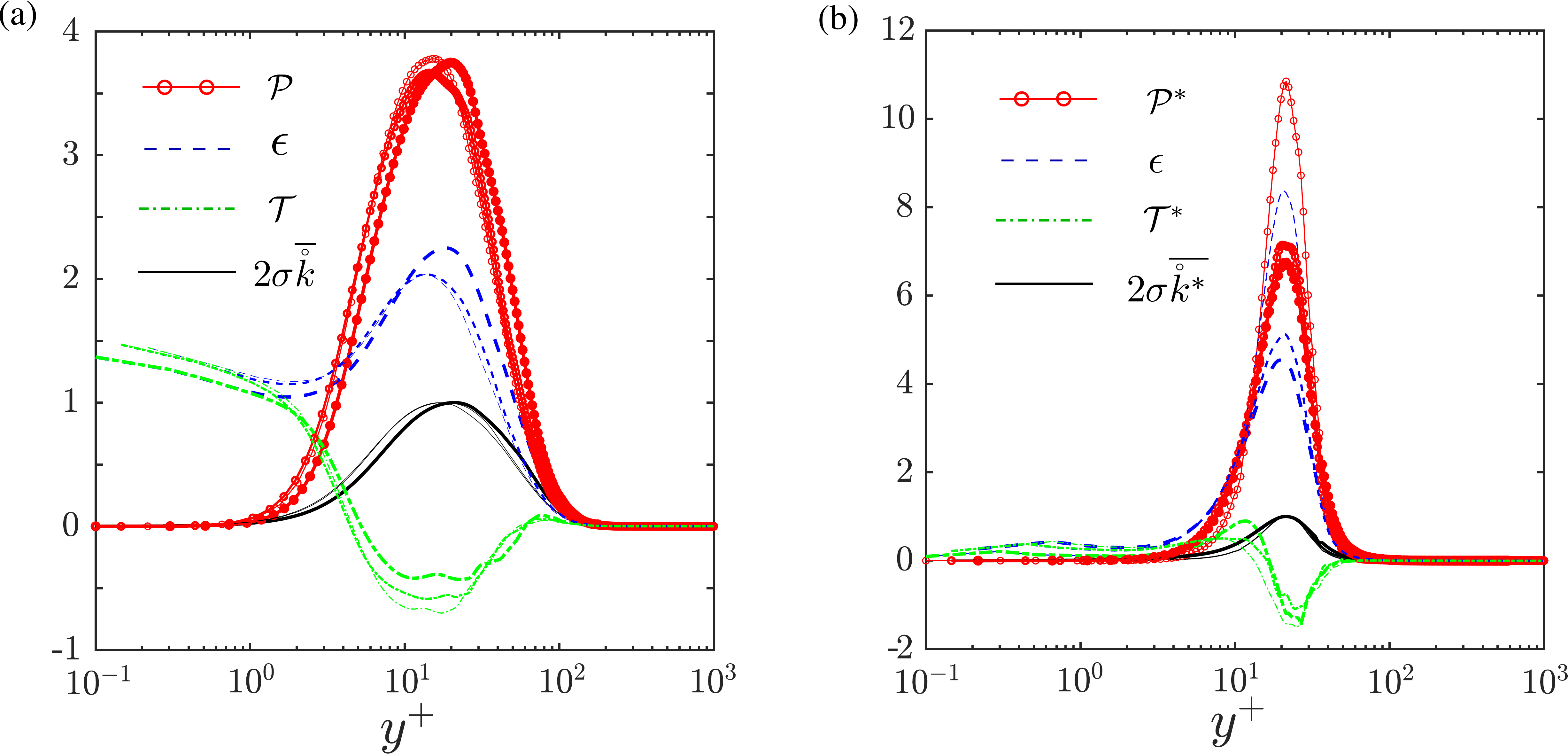}
    \caption{Kinetic energy budget for (a) the linearized forward and (b) the adjoint equations. Results for $Re_{\tau} = \{180, 590, 1000\}$ and shown with increasing line thickness. Curves are normalized such that $\max_y 2\sigma \overline{\mathring{k}} = \max_y 2\sigma \overline{\mathring{k}^{*}} = 1$.}
    \label{Fig:Budget}
\end{figure}

The various terms in the kinetic energy equations (\ref{Eq:tke_forward}) and (\ref{Eq:tke_adjoint}) are reported in figure \ref{Fig:Budget}.
For the forward flow, the profiles are qualitatively similar to the recent results by \citet{nikitin2018characteristics} at $Re_{\tau}$ around $390$;
In contrast to the forward terms, the adjoint profiles are confined to a narrower wall-normal extent $y^+ \le 60$, have more compact support and have a relatively more prominent peak in the buffer layer where the adjoint activity is most intense.
The cause of the difference between the forward and adjoint terms can perhaps be understood in terms of the reversal in the turbulent motions in the adjoint evolution: Sweeps that are prominent near the wall in the forward dynamics become ejections in the adjoint; and pronounced ejections above the buffer layer in the forward equations become sweeps in the adjoint.
Ultimately the very large concentrated production of the adjoint energy in the buffer layer is responsible for the exponential amplification in reverse time, which has the dual role of increasing the gradient of the cost function and rendering the state estimation problem ill-conditioned with expanding assimilation horizon; the buffer layer thus obscures the interpretability of wall observations.
However, when these results are viewed through the lens of the spectral analysis of the Hessian (\S\ref{Sec:EigenAnalysis}), it becomes evident that wall observations remain sensitive to low $k_x$ flow structures beyond the buffer layer, which have been related to outer-inner interactions in wall turbulence. 
Taken all together, the present statistical results and the Hessian analysis provide a comprehensive understanding of the interpretatbility of wall observations in the attempt to estimate the initial condition of channel flow turbulence. 

\section{Conclusion}
\label{Sec:Conclusion}

The nonlinear and chaotic natures of turbulence render its reconstruction from limited observations a challenging problem. 
The present work focused on the canonical configuration of turbulent channel flow.  
While all the vorticity in this flow has its origin at the wall, estimating the turbulent state from  observations of the wall shear stresses of wall pressure is notoriously difficult. 

The present effort focuses on the interpretability of the wall measurements, from the perspective of adjoint-variational data assimilation (4DVar).  In this approach, the estimation problem is formulated as a constrained optimization where we attempt to identify the initial condition whose nonlinear Navier-Stokes evolution reproduces all the available observations.  
Discrepancies between the model predictions and the available observations define the cost function to minimize, and which also features in the forcing term to the adjoint equations that are marched back in time.  The outcome of one forward-adjoint loop is the gradient of the cost function that is used in the gradient-based minimization procedure.  The accuracy of the state estimation depends on a number of factors, including the sensitivity of the observations to the flow state which is related ot the geometry of the cost function, most importantly its gradient and Hessian. In order to ensure accuracy of of our computations, a discrete adjoint is adopted which satisfies the forward-adjoint duality relation to machine precision.

In order to frame the discussion, we provided a summary of recent results by \citet{mengze2021} for estimating the initial state of turbulent channel flow from wall observations, at friction Reynolds numbers $Re_\tau = \{100, 180, 392, 590\}$. 
The assimilated states were obtained starting with an initial guess from a linear stochastic estimation and performing 100 iterations of the 4DVar algorithm, for each of the reported Reynolds numbers. 
The predicted flow state displayed important characteristics: 
(i) The near-wall turbulence was accurately reconstructed with a near-perfect correlation with the true flow that generated the observations.  This region therefore diminishes in physical height at higher Reynolds numbers. 
(ii) The correlation coefficient, however, decays precipitously across the buffer layer and into the channel core.  Despite the reduction in accuracy, the outer large-scale energetic motions which modulate the near-wall flow are captured by the estimation. 

The above characteristics were explained by examining the domain of dependence of wall observations and geometric properties of the state-estimation cost function for an instantaneous, isolated wall observation.
Specifically, we considered a scenario where the estimate of the initial flow field is infinitesimally close to the true solution and an instantaneous observation at $t=t_m$ is used to improve the estimate.  
At optimality the gradient of the cost function vanishes and its Hessian determines the behaviour of the optimization procedure.
In order to evaluate the Hessian, an efficient approach is introduced that exploits a forward-adjoint duality relation. 
The relevant adjoint field in this relation is initiated from an observation kernel at the wall; its spatio-temporal evolution in reverse time is the domain of dependence of the kernel, which has a universal behaviour over Reynolds numbers when scaled in viscous units; and the cross-correlation of this adjoint field at $\tau = t_m$ is the Hessian matrix.  
Ensemble averaging of the Hessian is performed over different observation locations $(x_m,z_m)$ in the wall plane and different starting times of adjoint marching. 
Additionally, periodicity in the horizontal plane enables decomposition of the Hessian into Fourier constituents in the spanwise and streamwise directions.
The eigenmodes of this representation were then analyzed to demonstrate the capacity to use wall observations for solving the state estimation problem.

When observing the streamwise wall shear-stress at time $t^+_m = 4$, the highest sensitivity to the initial state is to spanwise rolls with wavenumber pair $(k_x^+, k_z^+) \approx (0.12, 0)$; this component of the initial condition is therefore reliably targeted by the adjoint-variational state estimation.  For longer measurement times, similar spanwise rolls with smaller $k_x^+$ have the greatest influence on the measurement.
For most wavenumber pairs, the leading eigenvectors of the Hessian are concentrated near the wall ($y^+ \le 40$), and hence the turbulence is accurately estimated in this region.  
One important exception are modes with $(k_x^+,k_z^+) \approx (0,0.02)$ which remain finite outside the buffer layer, and which represent the sensitivity of wall measurements to outer large-scale structures.


The role of the observation time $t_m$ is noteworthy:  At small values of $t_m^+ \approx 4$, the Hessian eigenspectrum captures the sensitivity of wall observations to perturbations to the flow trajectory that may not have sufficient time to amplify or decay due to the dynamics of the forward operator.  In contrast, at larger values of $t_m^+ \approx 20$, the Hessian eigenspectrum shows higher sensitivity to lower wavenumbers because the associated perturbations to the flow trajectory can amplify in forward time and have a sizable impact on wall signature, while high-wavenumbers are dissipated by viscosity.  
This perspective was reinforced by projecting the Hessian onto the leading first and second modes from a proper orthogonal decomposition (POD) of channel flow turbulence. The projection yields the sensitivity of wall observations to the energetic POD structures.  The peak sensitivity of the original Hessian disappears in the projection because it was not associated with energetic flow structures but rather with near-wall small-scale turbulence which is the target of adjoint-assimilation at short times.  
At larger times $t^+_m = 20$, the contours of the eigenvalues of the original and projected Hessians are more aligned, both showing strong sensitivity of wall observations to streamwise-elongated and energetic flow structures.

The increase in observation time is accompanied by another important effect: the largest eigenvalues of the Hessian amplify exponentially and their separation from smaller ones increases, which renders the Hessian progressively more ill conditioned.
If observations are accumulated during a time horizon, late observations dominate the state estimation problem and the solution of the inverse problem becomes prone to errors.  Specifically, infinitesimal uncertainties in later observations can lead to large errors in the estimation of the initial state.

The starting estimate for practical data assimilation often deviates appreciably from the true flow state, and hence the gradient of the cost function as well as its Hessian are important. 
Both quantities are related to the adjoint field, and are affected by adjoint chaos at long reverse times which is markedly different from the more familiar forward counterpart. 
We contrasted the evolution of small-amplitude perturbations to the forward flow field and the adjoint model, using the linearized Navier-Stokes equations and their adjoint, respectively.  
The results provided a glimpse of the cause for the exponential growth of the adjoint in backward time. 
The energy of the adjoint variable peaks in the buffer layer, and its distribution has narrower support in the wall-normal direction than that of the forward variable.
By rescaling the energy equation by twice the Lyapunov amplification rate of perturbations, we are able to compute the different contributions to its budget.  
Production is concentrated in the buffer layer and decays quickly away from its peak, relative to the widely known profile for the forward problem. 
This high adjoint production and turbulent kinetic energy are the root of chaos in the buffer layer, which increases the slope of the cost function for long assimilation windows and render its Hessian ill-conditioned thus obscuring reconstruction of the initial state from wall data.  
In summary, while all the turbulence scales in the near-wall region can be precisely reconstructed, the buffer layer acts as a low-pass filter above which only low-frequency, large-scale, energetic motions can be accurately estimated from wall observations.


\par\bigskip
\noindent
\textbf{Declaration of interests.} 
The authors report no conflict of interest.
 

\appendix
\section{
Temporal behavior of the Hessian eigenmodes}
\label{Sec:Elevation}

The eigenmodes of the Hessian represent the sensitivity of the observations to perturbations to the flow state. 
For the majority of modes, the leading eigenfunction was concentrated in the near-wall region, expanded vertically with observation time, and decayed above the buffer layer (see figure \ref{Fig:ElevationMap2}).  
One important exception was noted for the modes with horizontal wavenumber vector $(k_x^+, k_z^+) \approx (0, 0.02)$ at $t_m^+ \ge 20$ that provide a connection between wall observations and the outer large-scale structures.  
In order to characterize the general wall-normal extent of penetration of the Hessian eigenmodes into the channel, we adopt two measures , $y_c$ and $y_{0.1}$. 
The first is the moment 
\begin{equation}
	y_c \equiv \frac{\displaystyle\int_0^1 y |\hat{\mathbf{u}}| dy}{\displaystyle\int_0^1 |\hat{\mathbf{u}}| dy},
\end{equation}
where $|\hat{\mathbf{u}}| \equiv \sqrt{|\hat{u}|^2+|\hat{w}|^2+|\hat{w}|^2}$, and the second is the height at which the eigenvector decays to ten percent of its maximum value,
\begin{equation}
    y_{0.1} = \max{ \left\{y \;  ~|~ \; \left({|\hat{\mathbf{u}}(y)|}/{|\hat{\mathbf{u}}|_{max}}\right) = 0.1\right\}}.
\end{equation}

\begin{figure}
    \centering
    \includegraphics[width = 0.8\textwidth]{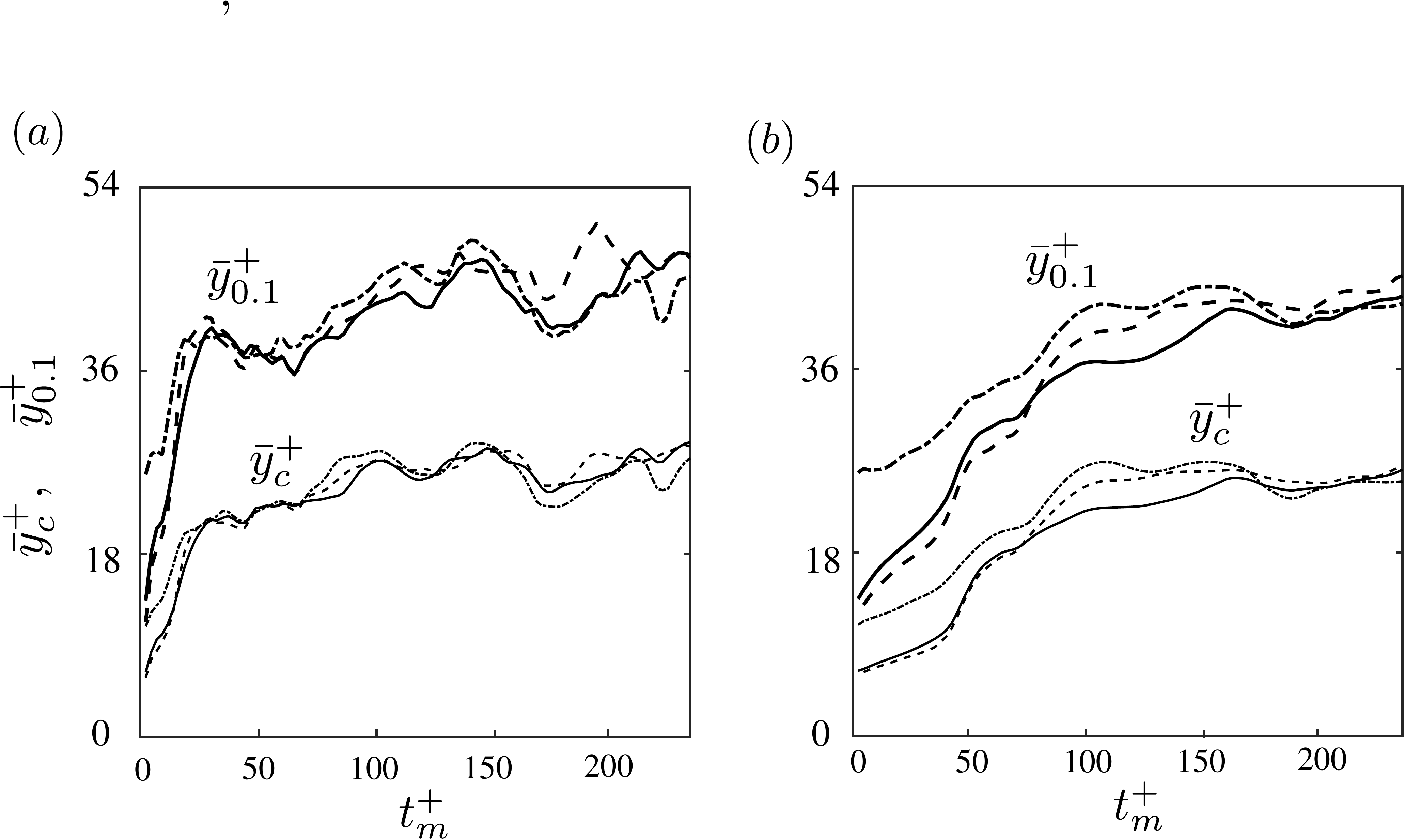}
    \caption{Measures of persistence of the Hessian eigenfunctions away from the wall, $\bar{y}_c$ and $\bar{y}_{0.1}$, averaged over the leading eigenvectors for every horizontal wavenumber pair. The Reynolds number is $Re_{\tau} = 180$. Results correspond to observing (\sampleline{})  $\partial U/\partial y \vert_{wall}$, (\sampleline{dashed})  $\partial W/\partial y \vert_{wall}$, and (\sampleline{dash pattern=on .7em off .2em on .2em off .2em}) $P \vert_{wall}$.
    (a) Observation at one time instance $t_m$; 
    (b) Observations at every time instance within the assimilation window $0 < t^{\prime} \le t_m$.
    }
    \label{Fig:yloc}
\end{figure}

\begin{figure}
    \centering
    \includegraphics[width = 0.5\textwidth]{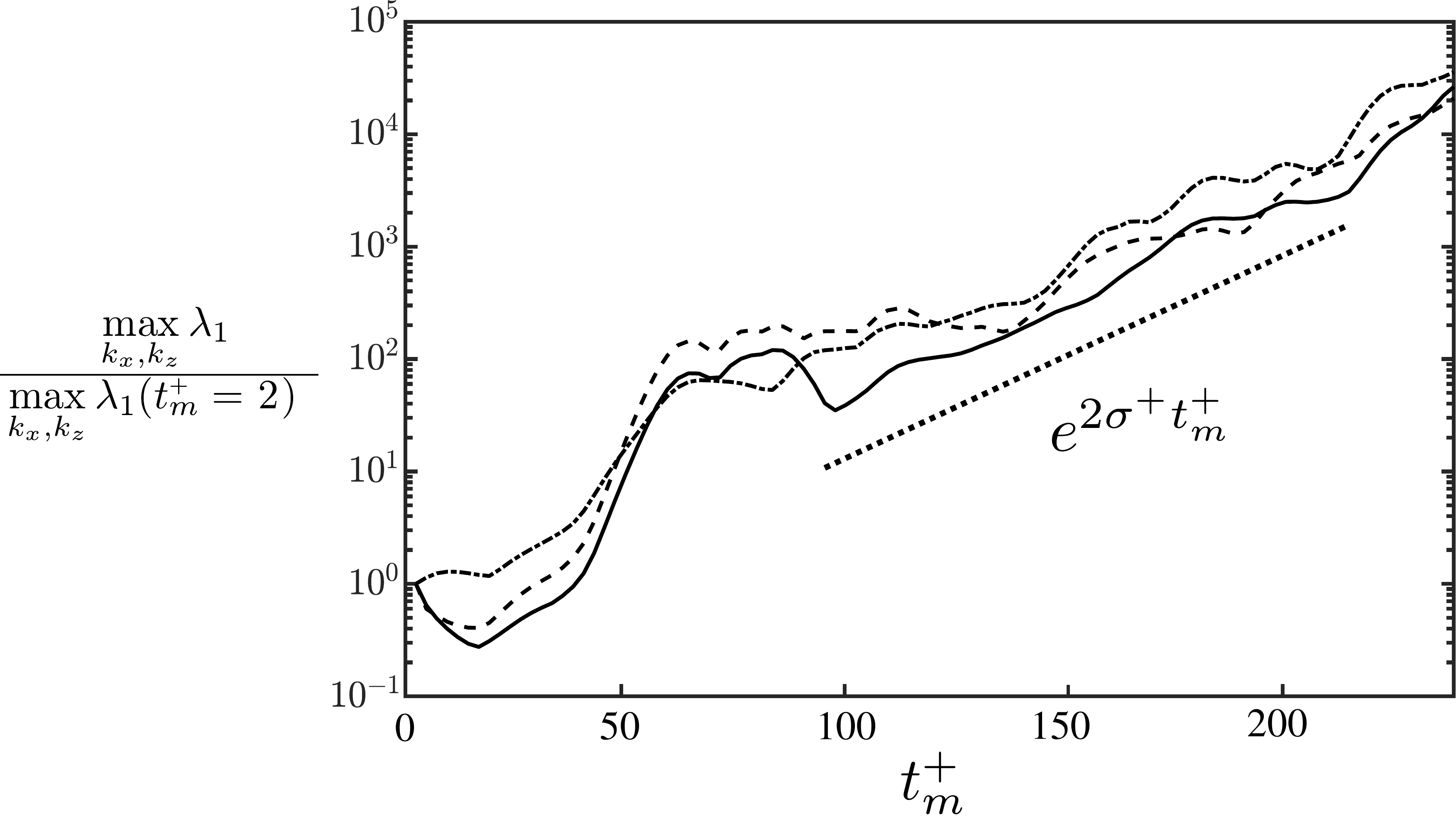}
    \caption{The supremum eigenvalue of the Hessian matrices associated with instantaneous observations at all horizontal wavenumber pairs $(k_x^+, k_z^+)$ as a function of the observation time $t_m^+$ for $Re_{\tau} = 180$. 
    Different wall-observation modalities are shown:  $\partial U/\partial y\big\rvert_{wall}$ (\sampleline{}) , $\partial W/\partial y\big\rvert_{wall}$ (\sampleline{dashed})  and $P\big\rvert_{wall}$ (\sampleline{dash pattern=on .7em off .2em on .2em off .2em}).
    Lyapunov amplification at $\sigma^+ = 0.0215$ is marked for comparison.}
    \label{Fig:MaxLambda}
\end{figure}

For Reynolds number $Re_{\tau} = 180$, we considered the Hessian eigenspectrum due to an instantaneous observation at time $t_m$, and averaged $y_c$ and $y_{0.1}$ over all horizontal wavenumber vectors, $\overline{y_c} \equiv \langle y_c\rangle_{k_xk_z}$ and $\overline{y_{0.1}} \equiv \langle y_{0.1}\rangle_{k_xk_z}$.  The results are shown as a function of $t_m^+$ in figure \ref{Fig:yloc}.  
While both measures initially increase with observation time, they level off beyond a few time units.  
The value of $\overline{y_c}$ never exceeds the buffer layer, while the less conservative estimate remains in the range $\overline{y_{0.1}^+} < 50$.  The elongated modes which penetrate beyond the buffer layer are therefore not reflected in the average, but are important to recall since they enable the reconstruction of the outer large-scale structures.  

Since in practice observations are often available not just at one time instance but rather over an assimilation time window, the cost function is integrated in time, i.e.\,all the available observations are used to estimate of the initial condition.
The relevant cost function is therefore the time integrated form. 
For example, when we observe $\partial U/\partial y \vert_{wall}$, the time-integrated cost function is,
\begin{equation}
\label{Eqn:CostFunction_timeintegration}
     J_{u} = \frac{1}{2S}\int_0^{t_m}\int_S \left( \frac{\partial u}{\partial y}\bigg\rvert_{(\mathbf{x}_m,t^{\prime})} \right)^2 dx_m dz_m dt^{\prime},
\end{equation}
and the associated Hessian involves the time integral, 
\begin{equation}
\label{Eqn:Hessian_timeIntegration}
    H_u = \frac{\partial^2 J_u}{\partial \mathbf{u}_0\partial \mathbf{u}_0} =\frac{1}{S} \int_0^{t_m}\int_S \mathbf{u}^*\left(\bullet;\mathbf{x}_m,t^{\prime}\right)\mathbf{u}^*\left(\bullet;\mathbf{x}_m,t^{\prime}\right) dx_m dz_m dt^{\prime},
\end{equation}
where $\bullet$ refers to $(\mathbf x, t=0)$. We can similarly evaluate $H_w$ and $H_p$.
In other words, the Hessian associated with estimating the initial state from the full time-history of observations is the time integral of the Hessian associated with isolated, or instantaneous, observations.  
The integral is dominated by late observations because the adjoint field amplifies exponentially with $t_m$. 
As such, we can anticipate that the results from figure \ref{Fig:yloc}$a$ remain qualitatively unchanged when observations are accumulated over the assimilation window.  
This expectation is verified in figure \ref{Fig:yloc}$b$ where $\overline{y_c}$ and $\overline{y_{0.1}}$ are plotted, evaluated from the eigenfunction of the time-integrated Hessian. 
These results supplement the discussion of figure \ref{Fig:ElevationMap2} in the main text, and provide statistical confirmation of the challenge of estimating channel-flow turbulence from wall observations.

The discussion at the end of \S\ref{Sec:EigenAnalysis} referred to the time-dependence of the supremum eigenvalue of the Hessian associated with instantaneous observations.  This dependence is reported in figure \ref{Fig:MaxLambda} as a function of $t_m^+$. 
After an initial transient, the eigenvalues grow exponentially due to the chaotic nature of the adjoint system.
This trend, and the larger separation between the supremum and infimum (figure \ref{Fig:ElevationMap2}) with observation time, demonstrate that the Hessian becomes increasingly ill-conditioned. As a result, the state-estimation problem becomes progressively more challenging to solve accurately.

\bibliographystyle{jfm}
\bibliography{Manuscript}

\begin{thebibliography}{44}
\expandafter\ifx\csname natexlab\endcsname\relax\def\natexlab#1{#1}\fi
\def\au#1{#1} \def\ed#1{#1} \def\yr#1{#1}\def\at#1{#1}\def\jt#1{\textit{#1}}
  \def\bt#1{#1}\def\bvol#1{\textbf{#1}} \def\vol#1{#1} \def\pg#1{#1}
  \def\publ#1{#1}\def\arxiv#1{#1}\def\org#1{#1}\def\st#1{\textit{#1}}

\bibitem[Abe {\em et~al.\/}(2004)Abe, Kawamura \& Choi]{abe2004very}
{\sc \au{Abe, Hiroyuki}, \au{Kawamura, Hiroshi} \& \au{Choi, Haecheon}}
  \yr{2004}  \at{Very large-scale structures and their effects on the wall
  shear-stress fluctuations in a turbulent channel flow up to re $\tau$= 640}.
  \jt{Journal of Fluids Engineering}  \bvol{126}~(5),  \pg{835--843}.

\bibitem[Adrian \& Moin(1988)]{Adrian1988LSE}
{\sc \au{Adrian, Ronald~J} \& \au{Moin, Parviz}} \yr{1988}  \at{Stochastic
  estimation of organized turbulent structure: homogeneous shear flow}.
  \jt{J.~Fluid Mech.}  \bvol{190},  \pg{531--559}.

\bibitem[Bewley \& Protas(2004)]{bewley2004skin}
{\sc \au{Bewley, Thomas~R} \& \au{Protas, Bartosz}} \yr{2004}  \at{Skin
  friction and pressure: the “footprints” of turbulence}.  \jt{Physica D:
  Nonlinear Phenomena}  \bvol{196}~(1-2),  \pg{28--44}.

\bibitem[Buchta \& Zaki(2021)]{buchta2021envar}
{\sc \au{Buchta, David~A.} \& \au{Zaki, Tamer~A.}} \yr{2021}
  \at{Observation-infused simulations of high-speed boundary-layer transition}.
   \jt{Journal of Fluid Mechanics}  \bvol{916},  \pg{A44}.

\bibitem[Cantwell {\em et~al.\/}(1978)Cantwell, Coles \&
  Dimotakis]{cantwell_1978}
{\sc \au{Cantwell, Brian}, \au{Coles, Donald} \& \au{Dimotakis, Paul}}
  \yr{1978}  \at{Structure and entrainment in the plane of symmetry of a
  turbulent spot}.  \jt{Journal of Fluid Mechanics}  \bvol{87}~(4),
  \pg{641–672}.

\bibitem[Chandramoorthy {\em et~al.\/}(2019)Chandramoorthy, Fernandez, Talnikar
  \& Wang]{chandramoorthy2019feasibility}
{\sc \au{Chandramoorthy, Nisha}, \au{Fernandez, Pablo}, \au{Talnikar,
  Chaitanya} \& \au{Wang, Qiqi}} \yr{2019}  \at{Feasibility analysis of
  ensemble sensitivity computation in turbulent flows}.  \jt{AIAA Journal}
  \bvol{57}~(10),  \pg{4514--4526}.

\bibitem[Chevalier {\em et~al.\/}(2006)Chevalier, H{\oe}pffner, Bewley \&
  Henningson]{chevalier2006state}
{\sc \au{Chevalier, Mattias}, \au{H{\oe}pffner, J{\'e}r{\^o}me}, \au{Bewley,
  Thomas~R} \& \au{Henningson, Dan~S}} \yr{2006}  \at{State estimation in
  wall-bounded flow systems. {P}art 2. {T}urbulent flows}.  \jt{Journal of
  Fluid Mechanics}  \bvol{552},  \pg{167--187}.

\bibitem[Colburn {\em et~al.\/}(2011)Colburn, Cessna \&
  Bewley]{colburn2011state}
{\sc \au{Colburn, CH}, \au{Cessna, JB} \& \au{Bewley, TR}} \yr{2011}  \at{State
  estimation in wall-bounded flow systems. {P}art 3. {T}he ensemble {K}alman
  filter}.  \jt{Journal of Fluid Mechanics}  \bvol{682},  \pg{289--303}.

\bibitem[Deissler(1986)]{Deissler1986chaotic}
{\sc \au{Deissler, RG}} \yr{1986}  \at{Is {N}avier--{S}tokes turbulence
  chaotic?}  \jt{Phys. Fluids}  \bvol{29}~(5),  \pg{1453--1457}.

\bibitem[Dimet \& Talagrand(1986)]{Dimet1986_4dvar}
{\sc \au{Dimet, FranÇois-Xavier~Le} \& \au{Talagrand, Olivier}} \yr{1986}
  \at{Variational algorithms for analysis and assimilation of meteorological
  observations: theoretical aspects}.  \jt{Tellus A: Dyn. Meteorol. Oceanogr.}
  \bvol{38}~(2),  \pg{97--110}.

\bibitem[Encinar \& Jim{\'e}nez(2019)]{encinar2019logarithmic}
{\sc \au{Encinar, Miguel~P} \& \au{Jim{\'e}nez, Javier}} \yr{2019}
  \at{Logarithmic-layer turbulence: A view from the wall}.  \jt{Physical Review
  Fluids}  \bvol{4}~(11),  \pg{114603}.

\bibitem[Eyink {\em et~al.\/}(2004)Eyink, Haine \& Lea]{eyink2004ruelle}
{\sc \au{Eyink, GL}, \au{Haine, TWN} \& \au{Lea, DJ}} \yr{2004}  \at{Ruelle's
  linear response formula, ensemble adjoint schemes and {L}{\'e}vy flights}.
  \jt{Nonlinearity}  \bvol{17}~(5),  \pg{1867}.

\bibitem[H{\oe}pffner {\em et~al.\/}(2005)H{\oe}pffner, Chevalier, Bewley \&
  Henningson]{hoepffner2005state}
{\sc \au{H{\oe}pffner, J{\'e}r{\^o}me}, \au{Chevalier, Mattias}, \au{Bewley,
  Thomas~R} \& \au{Henningson, Dan~S}} \yr{2005}  \at{State estimation in
  wall-bounded flow systems. {P}art 1. {P}erturbed laminar flows}.  \jt{Journal
  of Fluid Mechanics}  \bvol{534},  \pg{263--294}.

\bibitem[Hunt \& Durbin(1999)]{hunt1999perturbed}
{\sc \au{Hunt, JCR} \& \au{Durbin, PA}} \yr{1999}  \at{Perturbed vortical
  layers and shear sheltering}.  \jt{Fluid dynamics research}  \bvol{24}~(6),
  \pg{375}.

\bibitem[Hwang {\em et~al.\/}(2016)Hwang, Lee, Sung \& Zaki]{hwang2016inner}
{\sc \au{Hwang, Jinyul}, \au{Lee, Jin}, \au{Sung, Hyung~Jin} \& \au{Zaki,
  Tamer~A}} \yr{2016}  \at{Inner-outer interactions of large-scale structures
  in turbulent channel flow}.  \jt{Journal of Fluid Mechanics}  \bvol{790},
  \pg{128}.

\bibitem[Kalmikov \& Heimbach(2014)]{kalmikov2014hessian}
{\sc \au{Kalmikov, Alexander~G} \& \au{Heimbach, Patrick}} \yr{2014}  \at{A
  {H}essian-based method for uncertainty quantification in global ocean state
  estimation}.  \jt{SIAM Journal on Scientific Computing}  \bvol{36}~(5),
  \pg{S267--S295}.

\bibitem[Kerswell(2018)]{Kerswell2018nonlinear}
{\sc \au{Kerswell, R.~R.}} \yr{2018}  \at{Nonlinear nonmodal stability theory}.
   \jt{Annual Review of Fluid Mechanics}  \bvol{50},  \pg{319--345}.

\bibitem[Kleist \& Ide(2015{\natexlab{{\em a\/}}})]{kleist2015osse1}
{\sc \au{Kleist, Daryl~T} \& \au{Ide, Kayo}} \yr{2015{\natexlab{{\em a\/}}}}
  \at{An osse-based evaluation of hybrid variational--ensemble data
  assimilation for the {NCEP} {GFS}. {P}art {I}: System description and
  {3D}-hybrid results}.  \jt{Monthly Weather Review}  \bvol{143}~(2),
  \pg{433--451}.

\bibitem[Kleist \& Ide(2015{\natexlab{{\em b\/}}})]{kleist2015osse2}
{\sc \au{Kleist, Daryl~T} \& \au{Ide, Kayo}} \yr{2015{\natexlab{{\em b\/}}}}
  \at{An osse-based evaluation of hybrid variational--ensemble data
  assimilation for the {NCEP} {GFS}. {P}art {II}: {4DEnVar} and hybrid
  variants}.  \jt{Monthly Weather Review}  \bvol{143}~(2),  \pg{452--470}.

\bibitem[Lee {\em et~al.\/}(2017)Lee, Sung \& Zaki]{lee2017signature}
{\sc \au{Lee, Jin}, \au{Sung, Hyung~Jin} \& \au{Zaki, Tamer~A}} \yr{2017}
  \at{Signature of large-scale motions on turbulent/non-turbulent interface in
  boundary layers}.  \jt{Journal of Fluid Mechanics}  \bvol{819},
  \pg{165--187}.

\bibitem[Li {\em et~al.\/}(2020)Li, Zhang, Dong \& Abdullah]{Li2020}
{\sc \au{Li, Yi}, \au{Zhang, Jianlei}, \au{Dong, Gang} \& \au{Abdullah,
  Naseer~S}} \yr{2020}  \at{Small-scale reconstruction in three-dimensional
  kolmogorov flows using four-dimensional variational data assimilation}.
  \jt{Journal of Fluid Mechanics}  \bvol{885}.

\bibitem[Luchini \& Bottaro(2014)]{Luchini2014adjoint}
{\sc \au{Luchini, Paolo} \& \au{Bottaro, Alessandro}} \yr{2014}  \at{Adjoint
  equations in stability analysis}.  \jt{Annual Review of fluid mechanics}
  \bvol{46},  \pg{493--517}.

\bibitem[Lumley(1967)]{Lumley1967}
{\sc \au{Lumley, J.~L.}} \yr{1967}  \at{The structure of inhomogeneous
  turbulence}.  \bt{In {\em Atmospheric Turbulence and Wave Propagation\/} (ed.
  \ed{A.~M. Yaglom \& V.~I. Tatarski})}, ,  \vol{vol. 177},  \pg{pp. 166--178}.
   \publ{Nauka, Moscow}.

\bibitem[Marxen \& Zaki(2019)]{marxen2019turbulence}
{\sc \au{Marxen, Olaf} \& \au{Zaki, Tamer~A}} \yr{2019}  \at{Turbulence in
  intermittent transitional boundary layers and in turbulence spots}.
  \jt{Journal of Fluid Mechanics}  \bvol{860},  \pg{350--383}.

\bibitem[Mathis {\em et~al.\/}(2009)Mathis, Hutchins \&
  Marusic]{mathis2009large}
{\sc \au{Mathis, Romain}, \au{Hutchins, Nicholas} \& \au{Marusic, Ivan}}
  \yr{2009}  \at{Large-scale amplitude modulation of the small-scale structures
  in turbulent boundary layers} .

\bibitem[Moin \& Moser(1989)]{moin_moser_1989}
{\sc \au{Moin, Parviz} \& \au{Moser, Robert~D.}} \yr{1989}
  \at{Characteristic-eddy decomposition of turbulence in a channel}.
  \jt{Journal of Fluid Mechanics}  \bvol{200},  \pg{471–509}.

\bibitem[Mons {\em et~al.\/}(2016)Mons, Chassaing, Gomez \&
  Sagaut]{mons2016reconstruction}
{\sc \au{Mons, Vincent}, \au{Chassaing, J-C}, \au{Gomez, Thomas} \& \au{Sagaut,
  Pierre}} \yr{2016}  \at{Reconstruction of unsteady viscous flows using data
  assimilation schemes}.  \jt{Journal of Computational Physics}  \bvol{316},
  \pg{255--280}.

\bibitem[Mons {\em et~al.\/}(2019)Mons, Wang \& Zaki]{mons2019kriging}
{\sc \au{Mons, Vincent}, \au{Wang, Qi} \& \au{Zaki, Tamer~A}} \yr{2019}
  \at{Kriging-enhanced ensemble variational data assimilation for scalar-source
  identification in turbulent environments}.  \jt{Journal of Computational
  Physics}  \bvol{398},  \pg{108856}.

\bibitem[Nikitin(2018)]{nikitin2018characteristics}
{\sc \au{Nikitin, Nikolay}} \yr{2018}  \at{Characteristics of the leading
  {L}yapunov vector in a turbulent channel flow}.  \jt{Journal of Fluid
  Mechanics}  \bvol{849},  \pg{942--967}.

\bibitem[Nocedal(1980)]{LBFGS}
{\sc \au{Nocedal, J.}} \yr{1980}  \at{Updating {quasi-Newton} matrices with
  limited storage}.  \jt{Mathematics of Computation}  \bvol{35}~(151),
  \pg{773--782}.

\bibitem[Papadimitriou \& Giannakoglou(2008)]{papadimitriou2008direct}
{\sc \au{Papadimitriou, DI} \& \au{Giannakoglou, KC}} \yr{2008}  \at{Direct,
  adjoint and mixed approaches for the computation of {H}essian in airfoil
  design problems}.  \jt{International Journal for Numerical Methods in Fluids}
   \bvol{56}~(10),  \pg{1929--1943}.

\bibitem[Phillips(1969)]{Phillips1969}
{\sc \au{Phillips, O~M}} \yr{1969}  \at{Shear-flow turbulence}.  \jt{Annual
  Review of Fluid Mechanics}  \bvol{1}~(1),  \pg{245--264}.

\bibitem[Rosenfeld {\em et~al.\/}(1991)Rosenfeld, Kawak \&
  Vinokur]{Rosenfeld_1991jcp}
{\sc \au{Rosenfeld, M.}, \au{Kawak, D.} \& \au{Vinokur, M.}} \yr{1991}  \at{A
  fractional step solution method for the unsteady incompressible
  {N}avier-{S}tokes equations in generalized curvilinear coordinate systems}.
  \jt{Journal of Computational Physics}  \bvol{94},  \pg{102 -- 137}.

\bibitem[Schmid(2007)]{Schmid2007nonmodal}
{\sc \au{Schmid, Peter~J}} \yr{2007}  \at{Nonmodal stability theory}.
  \jt{Annual Review of Fluid Mechanics}  \bvol{39},  \pg{129--162}.

\bibitem[Suzuki(2012)]{Suzuki2012}
{\sc \au{Suzuki, Takao}} \yr{2012}  \at{Reduced-order {K}alman-filtered hybrid
  simulation combining particle tracking velocimetry and direct numerical
  simulation}.  \jt{J.~Fluid Mech.}  \bvol{709},  \pg{249–288}.

\bibitem[Suzuki \& Hasegawa(2017)]{hasegawa2016estimation}
{\sc \au{Suzuki, Takao} \& \au{Hasegawa, Yosuke}} \yr{2017}  \at{Estimation of
  turbulent channel flow at {$Re_{\tau}=100$} based on the wall measurement
  using a simple sequential approach}.  \jt{Journal of Fluid Mechanics}
  \bvol{830},  \pg{760--796}.

\bibitem[Taira {\em et~al.\/}(2017)Taira, Brunton, Dawson, Rowley, Colonius,
  McKeon, Schmidt, Gordeyev, Theofilis \& Ukeiley]{Taira2017modal}
{\sc \au{Taira, Kunihiko}, \au{Brunton, Steven~L}, \au{Dawson, Scott~TM},
  \au{Rowley, Clarence~W}, \au{Colonius, Tim}, \au{McKeon, Beverley~J},
  \au{Schmidt, Oliver~T}, \au{Gordeyev, Stanislav}, \au{Theofilis, Vassilios}
  \& \au{Ukeiley, Lawrence~S}} \yr{2017}  \at{Modal analysis of fluid flows: An
  overview}.  \jt{AIAA Journal}  \bvol{55}~(12),  \pg{4013--4041}.

\bibitem[Vishnampet {\em et~al.\/}(2015)Vishnampet, Bodony \&
  Freund]{Vishnampet2015}
{\sc \au{Vishnampet, Ramanathan}, \au{Bodony, Daniel~J} \& \au{Freund,
  Jonathan~B}} \yr{2015}  \at{A practical discrete-adjoint method for
  high-fidelity compressible turbulence simulations}.  \jt{Journal of
  Computational Physics}  \bvol{285},  \pg{173--192}.

\bibitem[Wang {\em et~al.\/}(2019{\natexlab{{\em a\/}}})Wang, Wang \&
  Zaki]{mengze2019discrete}
{\sc \au{Wang, Mengze}, \au{Wang, Qi} \& \au{Zaki, Tamer~A.}}
  \yr{2019{\natexlab{{\em a\/}}}}  \at{Discrete adjoint of fractional-step
  incompressible {N}avier-{S}tokes solver in curvilinear coordinates and
  application to data assimilation}.  \jt{Journal of Computational Physics} .

\bibitem[Wang \& Zaki(2021)]{mengze2021}
{\sc \au{Wang, Mengze} \& \au{Zaki, Tamer~A}} \yr{2021}  \at{State estimation
  in turbulent channel flow from limited observations}.  \jt{Journal of Fluid
  Mechanics}  \bvol{917},  \pg{A9}.

\bibitem[Wang {\em et~al.\/}(2019{\natexlab{{\em b\/}}})Wang, Hasegawa \&
  Zaki]{wang_hasegawa_zaki_2019}
{\sc \au{Wang, Qi}, \au{Hasegawa, Yosuke} \& \au{Zaki, Tamer~A.}}
  \yr{2019{\natexlab{{\em b\/}}}}  \at{Spatial reconstruction of steady scalar
  sources from remote measurements in turbulent flow}.  \jt{Journal of Fluid
  Mechanics}  \bvol{870},  \pg{316–352}.

\bibitem[You \& Zaki(2019)]{you2019tbl}
{\sc \au{You, Jiho} \& \au{Zaki, Tamer~A.}} \yr{2019}  \at{Conditional
  statistics and flow structures in turbulent boundary layers buffeted by
  free-stream disturbances}.  \jt{Journal of Fluid Mechanics}  \bvol{866},
  \pg{526–566}.

\bibitem[Zaki(2013)]{zaki2013streaks}
{\sc \au{Zaki, Tamer~A}} \yr{2013}  \at{From streaks to spots and on to
  turbulence: exploring the dynamics of boundary layer transition}.  \jt{Flow,
  turbulence and combustion}  \bvol{91}~(3),  \pg{451--473}.

\bibitem[Zaki \& Saha(2009)]{zaki2009shear}
{\sc \au{Zaki, Tamer~A} \& \au{Saha, Sandeep}} \yr{2009}  \at{On shear
  sheltering and the structure of vortical modes in single-and two-fluid
  boundary layers}.  \jt{Journal of Fluid Mechanics}  \bvol{626},
  \pg{111--147}.

\end{thebibliography}

\end{document}